\documentclass[pre,reprint,twocolumn,noshowpacs]{revtex4}

\usepackage{epsfig}
\usepackage[]{natbib}
\usepackage{color}
\usepackage{amsmath}
\usepackage[toc]{appendix}

\usepackage{amsfonts} 
\usepackage{color} 
\usepackage{hyperref}
\usepackage{graphicx,amssymb}

\begin{document}
\title{
Conditional $1/f^{\alpha}$ noise: from single molecules to macroscopic measurements }

\author{N. Leibovich}
\author{E. Barkai}
\affiliation{Department of Physics, Institute of Nanotechnology and Advanced Materials, Bar-Ilan University, Ramat-Gan
5290002, Israel}

\begin{abstract}
We demonstrate that the measurement of  $1/f^{\alpha}$ noise at the single molecule or nano-object limit is remarkably distinct from the macroscopic measurement over a large sample. The single 	particle measurements yield a conditional time-dependent spectrum. 
However, the number of units fluctuating on the time scale of the experiment is increasing in such a way that the macroscopic measurements appear perfectly  stationary.
The single particle power spectrum is a conditional spectrum, in the sense that we must make a distinction between idler and non-idler units on the time scale
of the experiment. 
We demonstrate our results based on stochastic and deterministic models, in particular the well known superposition of Lorentzians approach, the blinking quantum dot model,
and deterministic dynamics generated by non-linear mapping. Our results show that the $1/f^\alpha$ spectrum is inherently nonstationary even if the macroscopic measurement completely obscures the underlying time dependence of the phenomena. 
\end{abstract}
\maketitle
%
\section{introduction}
In many experiments the measured power spectral density is 
\begin{equation}
S(\omega) \propto \omega^{-\alpha}
\label{eq:01}
\end{equation}
with $0.5<\alpha<1.5$. 
This behavior is practically universal as it is  found in a wide range of systems ranging from electronic devices, 
geological data, blinking quantum dots, and currents in ion channels to name only a few examples \cite{VanDerZiel,
Dutta,Keshner,Weissman,Hooge81,silvestri2009,herault2015,Sadegh,Caloyannides,Mandelbrot69,solo}. 
Such $1/f^{\alpha}$ fluctuations are found down to the lowest frequencies measured
which are of the order of $2 \pi/t$, where $t$ is the measurement time. 
For example $t$ is roughly an hour for blinking quantum dots \cite{Sadegh}, three months for careful measurements of voltage fluctuations in semiconductors \cite{Caloyannides} or years for geological data \cite{Mandelbrot69}. 
 
Many papers and reviews, by careful analysis of macroscopic data, propagate the idea that the $1/f^{\alpha}$ phenomenon is based on standard concepts of stationarity  \cite{Dutta,Weissman,Hooge81}, ``in the absence of overwhelming evidence 
to the contrary''  \cite{Dutta}.  This has a vast consequence, since stationarity implies  the standard definition of the spectrum and its connection to the underlying stationary correlation function through the Wiener-Khinchin theorem holds \cite{kubo2012statistical}. 

The $1/f^{\alpha}$ spectra,  Eq.~\eqref{eq:01}, are problematic since when $\alpha\ge 1$, the integral over the spectral density which gives the total power of the system, $\int_{1/t}^{\infty}\omega^{-\alpha}{\rm d}\omega$,  diverges when the measurement time goes to infinity due to the low-frequency behavior. 
However clearly for a bounded process the total power must be finite  $\int_0^{\infty}S(\omega){\rm d}\omega <\infty$ \cite{kubo2012statistical}. The demand for finite total power and the measurements of $1/f^{\alpha}$ noise in a vast array of systems seems to contradict each other \cite{VanDerZiel,Weissman,Dutta,Hooge81}.

One way to resolve this low-frequency paradox is to assume that the underlying process is nonstationary 
\cite{Mandelbrot1967,Niemann,watkins2016mandelbrot,LeibovichLong}.
Mandelbrot suggested that $1/f^{\alpha}$ power spectrum {\em ages} which means that
$S_t (\omega) \propto\omega^{-2+\beta} t^{-1+\beta}$,  so  $\alpha=2-\beta$  \cite{Mandelbrot1967}. Importantly, here the spectrum depends on the measurement time $t$ (see details below), and the total power remains finite  $\int_{1/t} ^\infty  S_t(\omega) {\rm d} \omega=\mbox{const}$ \cite{Niemann,LeibovichLong}. 
The time dependent amplitude
of $ S_t(\omega) $ provides  a normalizable spectral density,
therefore it should naturally appear in a bounded process.
In this scenario the spectrum is a density, as it should be,  in the sense that $S_t(\omega)$ is normalizable \cite{LeibovichLong}. Models of such nonstationary behavior are found in the theory of glasses \cite{Bouchaud,Crisanti}, blinking quantum dots, analytically and experimentally  \cite{Niemann, Sadegh}, nanoscale electrodes \cite{Krapf}, and interface fluctuations in the (1+1)-dimensional KPZ class, both experimentally and numerically, using liquid-crystal turbulence 
\cite{takeuchi2016}. 
Thus one school of thought supports the idea that the sample spectrum exhibits nonstationary features of a particular kind \cite{LeibovichPRL,DechantPRL,LeibovichLong,Rodriguez2015,
Krapf}. However, the others argue that while Mandelbrot's nonstationarity scenario is theoretically
elegant, it is not a universal explanation since it is backed only by several experiments \cite{Krapf,Sadegh, takeuchi2016}, and the spectrum is stationary \cite{Dutta,Weissman,Hooge81}. 

Why, 50 years after Mandelbrot pointed out the idea of an aged spectrum, is there only a few experimental evidences for a nonstationary power spectrum?  
In particular why do many measurements of $1/f^{\alpha}$ noise in condensed matter physics seemingly support the stationarity scenario? 
The key issue is the difference between macroscopic and microscopic measurements. 
A macroscopic measurement contains many microscopic realizations. For example consider a current, $I(t)$, flowing through a disordered  medium. The macroscopic system has many channels of current in it, distributed in a complicated way in the sample. The macroscopic measurement of the power spectrum corresponds to the total signal $I(t)$ generated in the sample, e.g. the spectrum of the total current. By microscopic measurements we mean local observations of individual units, e.g. the internal channels of current in the medium. Of course the signals from all those units, added together, yields the macroscopic measurement.

Following Mandelbrot \cite{Mandelbrot1967} we consider {\em conditional} measurements, which are important in the context of measuring of noise in the microscopic approach.  We denote the currents of units in the sample with  $I_j(t)$ where  $j=1,\ldots,N$ is the unit's index. The core of the concept is to separate the set $\{I_j(t)\}_{j=1}^N$  into two subsets; one with the realizations $I_j(t)$ which  appear stationary on the measurement time interval (this set is called ${\rm B}$).
The other set, the complementary one, contains all the other realizations (the set ${\rm B}^c$).
Of course to  distinguish between the two subsets one needs to be able to perform measurements one nano-object at a time, namely a microscopic measurements. For example a local observation of current in one small junction in the system.   Traditional spectral theory, based on Wiener-Khinchin theorem, holds for the stationary realizations in subset B. 
For a single particle measurement we consider a conditional measurement which is observed only from realizations which posses stationarity in the measurement time interval, then we average over the set B (i.e. averaging over the measured realizations set).  Note that the size of the set B depends on the measurement time; as we increase the measurement time the number of realizations in B is changing. Hence, a conditional spectrum, averaged over the set B, may depend on time, as we show in detail below.


Our goal in this paper is to show that there exists a profound difference between measurements of $1/f^{\alpha}$ noise on the single particle level, if compared with macroscopic measurements (defined below). As we will show, on the microscopic level, where one conditionally measures single particles, the power spectrum ages. However, macroscopic measurements yield a time independent spectrum. In that sense the tension between the  two conflicting approaches to $1/f^{\alpha}$ noise, i.e. the stationary versus the nonstationary communities, is reduced. 
As we show below,  aged spectrum is valid even for the most basic model of $1/f^\alpha$ noise, namely the distributed kinetic models for a variety of processes, e.g. two-state model, Ornstein-Uhlenbeck etc, which is partially analyzed in many review articles in the field. 

To demonstrate the wide validity of the main results we consider two classes of models.
We begin with the widely popular  distributed kinetic approach. Here, at least in principle, if one measures in a long time interval, the processes are stationary. In the second part of the paper we consider a very different class of processes, which are inherently nonstationary. We investigate stochastic models of blinking quantum dots, and a deterministic model of intermittency. 
While the two classes of models are vastly different the main conclusion is the same: there exist an essential difference between single particle and macroscopic measurements.


\section{ macroscopic  versus single-particle measurements }
\label{2Types}
Consider a large set of $N$ independent  processes $\{ I_j(t')\} $ observed in the time interval $[0,t]$ where $j\in \{1,2,...N\}$ is the unit's label. 
The single-particle spectrum is given by the periodogram \begin{equation}
S_{j}(\omega,t) =\frac{1}{t}\left|\int_0 ^t I_j(t') \exp(-i \omega t') {\rm d} t'\right|^2
\label{eq:SampleSpectrum}
\end{equation}
where $t$ is assumed to be long. For a stationary process this sample spectrum, Eq.~\eqref{eq:SampleSpectrum}, is given by Fourier transform of the autocorrelation function of the observable $I_j(t)$ via the Wiener-Khinchim theorem \cite{kubo2012statistical}. Additional smoothing of the sample spectrum is also routinely performed \cite{Press1992}, see also \cite{dean2016,Niemann}. 
In single-particle measurements one samples $n_s$ trajectories, i.e. $I_j(t')$ where $j=1 \ldots n_s$,
and then defines an average with respect to the measured processes, namely 
\begin{equation}
\langle S(\omega, t)\rangle_{{\rm sp}} = \sum_{k=1} ^{n_s} S_k (\omega, t)/n_s,
\label{eq:MicroSpectrum}
\end{equation}
 where $1\ll n_s\ll N$. 
Here $\langle . \rangle_{\rm sp}$ stands for single-particle measurements with ensemble averaging over a sub-set with size $n_s$. For a macroscopic measurement 
the spectrum of many independent processes, all measured in parallel, is 
\begin{equation}
{\cal S}(\omega , t)_{{\rm mac}}  = \sum_{j=1} ^N S_j (\omega, t),
\label{eq:MacroSpectrum}
\end{equation}
 (see also in App.~\ref{Paralleled}). Only if all the processes are identical and stationary we find that the macroscopic measurement is simply related to the single particle procedure via ${\cal S} (\omega)_{\rm mac} = N \langle S(\omega)\rangle _{{\rm sp}} $. Our goal is to show that this time-independent relation does not hold for models of $1/f^{\alpha}$ noise.  This is related to the way experimentalist choose the sub-ensemble of single particle measurements, as we now demonstrate with a simple two state model.


\section{Random Telegraph signal}
\label{RTN}
Consider a two-state telegraph process, where $I_j(t) = I_0$ or $I_j(t)=-I_0$ with sojourn times in each state, $\{{\cal T}^j_1,{\cal T}^j_2, \ldots \}$, that are exponentially distributed with mean $\tau_j$. After each waiting time the realization switches to the other state. 
For a long measurement time the process is stationary and ergodic such that \cite{Godreche}
\begin{equation}
\langle I_j(t_0) I_j(t_0+t')\rangle =I_0^2 \exp(-2t'/\tau_j),
\label{eq:03}
\end{equation} 
hence, using the Wiener-Khinchin theorem, the spectrum is \begin{equation}S_j(\omega )= I_0^2\frac{4\tau_j}{4+\omega^2\tau_j^2}.
\end{equation}
The value of $\tau_j$ varies from one 
molecule to the other. It is a quenched random variable in the sense that it is fixed for each process $I_j(t)$. 
This is a crude model for a single molecule in low temperature glasses, e.g. see \cite{boiron1999spectral,geva1997}.     

Consider a set of $N$ telegraph processes where the characteristic time scale $\tau_j $ is varying from one molecule to another, with a common probability density function (PDF) 
\begin{equation}
P(\tau)= {\cal N}\tau^{-\beta},
\label{PDF}
\end{equation}
with $0<\beta<1$. For such a distribution to be meaningful we introduce an upper and a lower cutoff, $\tau\in [\tau_{\rm min},\tau_{\rm max}]$, thus the normalization constant is ${\cal N}=(1-\beta)/[(\tau_{\rm max})^{1-\beta}-(\tau_{\rm min})^{1-\beta}]$.

The model of superimposed Lorentzian-shaped spectra with heavy-tailed distributed characteristic time $\tau$ is considered one of the best known explanations to the $1/f^{\alpha}$ phenomena \cite{McWhorter,Dutta,Hooge81,VanDerZiel,bernamont}. It was firstly developed nearly eight decades ago for vacuum tubes \cite{bernamont}, and later in the middle of the 50'th for semiconductors \cite{McWhorter}. See further discussion in Sec.~\ref{discussion}.

Now assume that a realization $j$ did not move at all on the time scale of the experiment, namely it is localized in its initial state during the entire measurement period, e.g. a unit with $\tau_j\gg t$. This noiseless unit does not contribute to the spectrum, namely its sample spectrum vanishes at natural frequencies $\omega=2\pi n/t$ when $n$ is a positive integer. Since units with no activity are not detectable, i.e. they are noiseless, experimentalists measure only the active units' subensemble. Hence the single particle spectrum is a conditional measurement.

\begin{figure}
	\centering
		\includegraphics[trim= 30 0 50 0, width=\columnwidth]{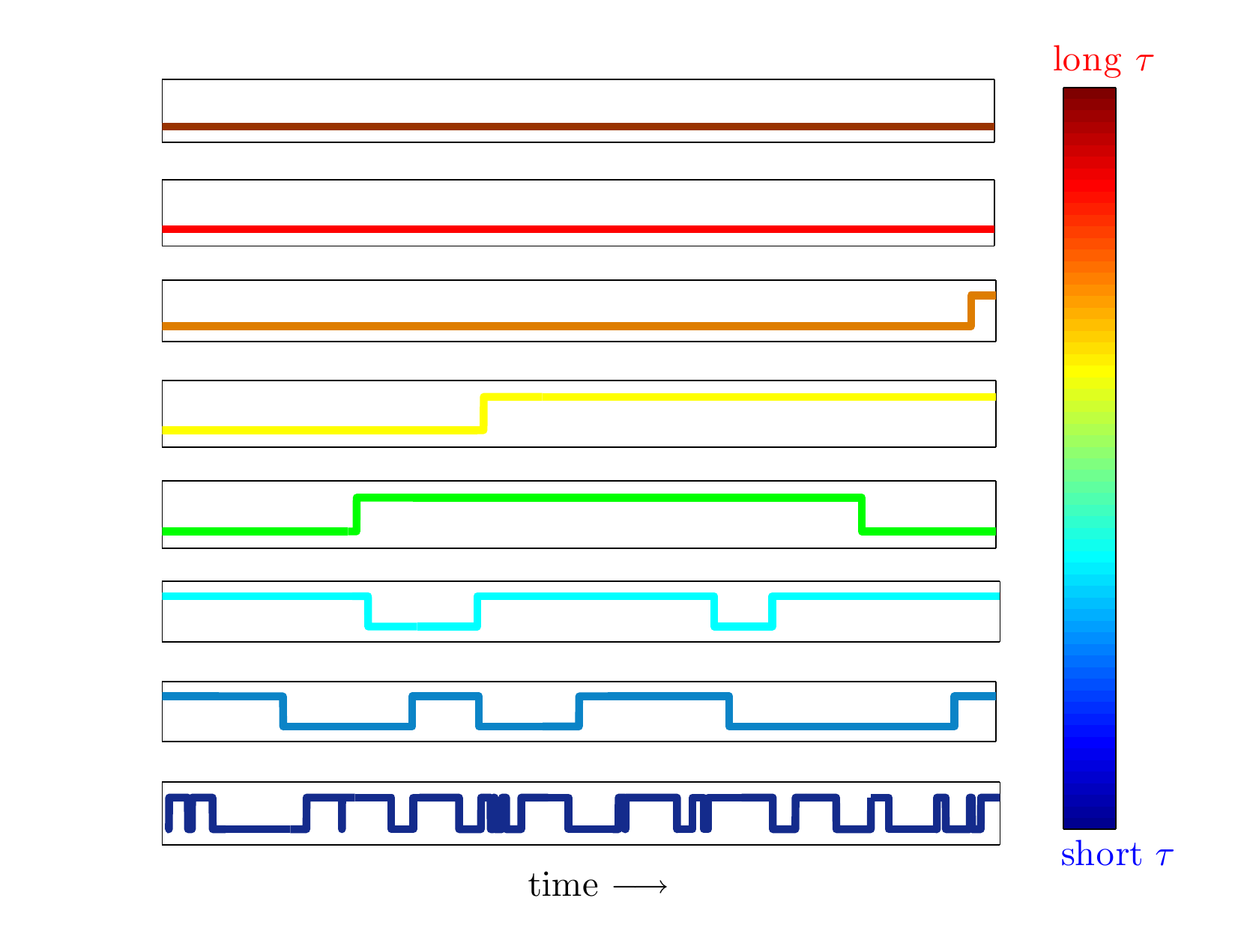}
        \caption{An illustration for eight dichotomous Poisson processes with a characteristic time scale $\tau_j$ varying from one realization to another. The realization with the longest $\tau$ is given at the top (dark red), then $\tau$ changes gradually where the shortest $\tau$ realization is given at the bottom (dark blue). It is clear that while changing the measurement time the size of the set of moving realizations changes as well.   }
	\label{fig:RTNExapmle}
\end{figure}

The probability of a realization with a given relaxation time $\tau$ to move in the time interval $[0,t]$ is 
\begin{equation}{\rm P}_0^{\rm mov}(t|\tau)=1-\exp(-t/\tau),
\label{eq:Condition}
\end{equation}
which is equivalent to the probability that the first sojourn time in the initial state is longer than the measurement time, i.e. ${\cal T}_1<t$, see Fig.~\ref{fig:RTNExapmle}.

Now we sample the spectrum of moving realizations only, thus we defined a conditional measurement of spectrum.  This protocol leave us with a subset of $\{\tau_j\}$ of the moving objects and the ensemble averaging is taken with respect to this subset. The normalized distribution of $\tau$ for the moving realization subset $\{\tau_j\}$ is $P(\tau){\rm P}_0^{\rm mov}\left(t|\tau\right) {\cal N}_t$, where the time-dependent normalization constant is given by
\begin{eqnarray} 
{\cal N}_t^{-1}&=& \int_{\tau_{\rm min}}^{\tau_{\rm max}}{\rm P}_0^{\rm mov}(t|\tau)P(\tau){\rm d}\tau = \label{eq:09}\\ \nonumber
&=&\int_{\tau_{\rm min}}^{\tau_{\rm max}}\left(1-e^{-t/\tau}\right)\frac{(1-\beta)\tau^{-\beta}}{\tau_{\rm max}^{1-\beta}-\tau_{\rm min}^{1-\beta}}{\rm d}\tau  \\ &\approx& \Gamma(\beta)\left(\frac{t}{\tau_{\rm max}}\right)^{1-\beta},
\nonumber 
\end{eqnarray}
where the limit $\tau_{\rm min}\ll t\ll \tau_{\rm max}$ was taken.
Averaging over the active particles' spectra yields 
\begin{eqnarray}
&&\langle S_t(\omega) \rangle_{{\rm sp}} \sim \nonumber\\ && {\cal N}_t I_0^2\int_{\tau_{\rm min}}^{\tau_{\rm max}}\frac{4\tau}{4+\omega^2\tau^2}{\rm P}_0^{\rm mov}(t|\tau)P(\tau){\rm d}(\tau)=
\label{eq:05}
\\ \nonumber &&{\cal N}_t I_0^2\int_{\tau_{\rm min}}^{\tau_{\rm max}}\frac{4\tau}{4+\omega^2\tau^2}\left(1-e^{-t/\tau}\right)\frac{(1-\beta)\tau^{-\beta}}{\tau_{\rm max}^{1-\beta}-\tau_{\rm min}^{1-\beta}}{\rm d}\tau.
\end{eqnarray}
In the limit $\tau_{\rm min}\ll t\ll \tau_{\rm max}$ we approximate the integration interval to $[0,\infty)$, then we find using Mathematica
\begin{eqnarray}
&&\int_{0}^{\infty}\frac{4\tau}{4+\omega^2\tau^2}\left(1-e^{-t/\tau}\right)\tau^{-\beta}{\rm d}\tau =  \\ \nonumber && = -t^{2-\beta} \Gamma (\beta-2) \,
   _1F_2\left[1;\frac{3}{2}-\frac{\beta}{2},2-\frac{\beta}{2};-\frac{1}{16} t^2
   \omega ^2 \right] \\ &&
   -\pi  2^{1-\beta} \omega^{\beta-2} \csc \left(\frac{\pi\beta}{2}\right) \left[\sec
   \left(\frac{\pi  \beta}{2}\right) \cos \left(\frac{\pi \beta+t   \omega}{2} \right)-1\right] \nonumber
\end{eqnarray}
where $_1F_2[a,b_1,b_2,x]$ refers to the Hypergeometric function. Then with the limit $\omega t \gg 1$ we obtain the single particle spectrum, conditioned on measurements of the moving processes
\begin{equation}
\langle S_t(\omega) \rangle_{{\rm sp}}\simeq I_0^2 A_{\beta}  \omega^{- 2 + \beta} t^{-1+\beta}
\label{eq03}
\end{equation}
with $A_\beta=2^{1-\beta}(1-\beta)\pi\csc\left(\frac{\pi \beta}{2}\right)/\Gamma(\beta)$ for $0<\beta<1$.
The conditional spectrum Eq.~\eqref{eq03} thus provides the averaged spectra per contributing unit. 

As was mentioned, from all $N$ units only a fraction of them  are contributing to the spectrum. 
The number of movers is $N_t=N\times \Gamma(\beta)(t /\tau_{\rm max})^{1-\beta}$ when $\tau_{\rm min}\ll t\ll\tau_{\rm max}$. 
The macroscopic measurement hence is 
\begin{equation}
{\cal S}(\omega)_{{\rm mac}} = N_t  \langle S(\omega, t) \rangle_{{\rm sp}}.
\label{eq04}
\end{equation}
The number of movers is increasing like $t^{1-\beta}$ while the spectrum  $\langle S(\omega,t)\rangle_{{\rm sp}}$ Eq.~\eqref{eq03} is decreasing as $t^{\beta-1}$  and we get from Eq.~\eqref{eq04}
 a macroscopic spectrum
\begin{equation}
{\cal S}(\omega)_{{\rm mac}} \simeq N(I_0)^2   
B_{\beta} \omega^{- 2 + \beta} ({\tau_{\rm max}})^{\beta-1}.
\label{eq05}
\end{equation}
with $B_\beta= 2^{1-\beta}(1-\beta)\pi\csc\left(\frac{\pi \beta}{2}\right)$. This spectrum is found in a range of frequencies as low as $1/t$: there is no flattening effect, and the macroscopic measurement appears stationary since it is measurement-time independent. 

The macroscopic noise in Eq.~\eqref{eq05}
is proportional to $N$ (as expected) multiplied by $(\tau_{\max})^{\beta-1}$, 
so unless one knows $N$ (which includes also noiseless idlers) he cannot determine the upper cutoff time $\tau_{\rm max}$ which remains nondetectable as long as it is much larger than~$t$.
We comment that while both $N$ and $\tau_{\rm max}$ are nondetectable, $N (\tau_{\rm max})^{\beta-1}$ is a measurable quantity since the number of movers is $N_t\propto (t/\tilde{\tau}_N)^{1-\beta}$ where $\tilde{\tau}_N=\tau_{\rm max}N^{1/(\beta-1)}$ may, in principle, be measured. 

\begin{figure}
	\centering
		\includegraphics[width=0.95\columnwidth, trim= 0 20 0 25]{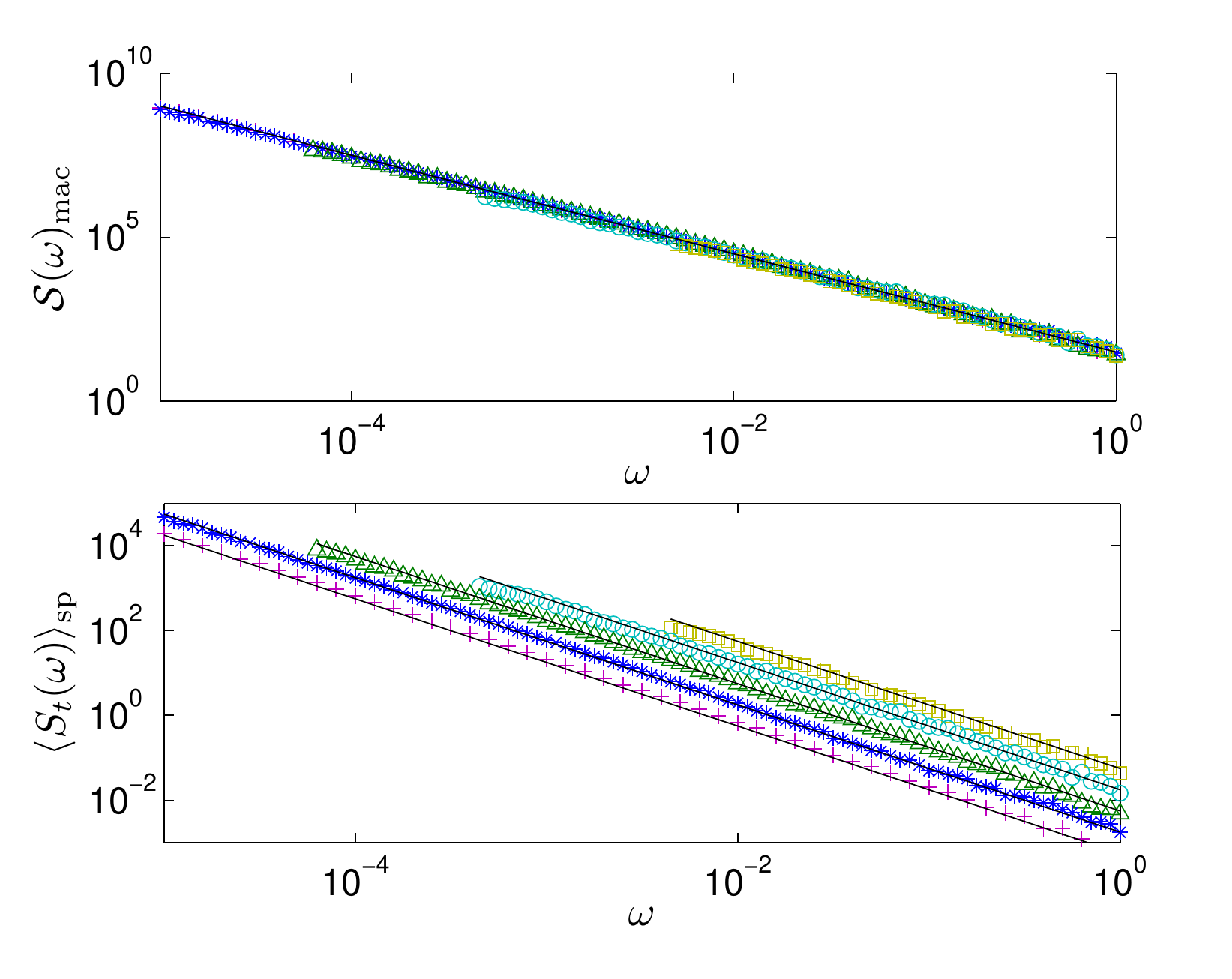}
		\caption{Simulation results for the macroscopic spectrum (upper panel) and the single particle conditional spectrum (lower). We use $N=10^5$ particles all following the two-state telegraph process with $I_0=1$ and relaxation times $\{\tau_j\}$ drawn from the fat-tailed PDF with $\beta=1/2$,  $\tau_{{\rm min}}=1$, $\tau_{{\max}} =10^8$ and $K=0$. The spectrum was measured at measurement times; $t=10^3$ (yellow squares), $t=10^4$ (cyan circles), $t=10^5$ (green triangles), $t=10^6$ (blue stars), and $t=10^7$ (pink crosses). The lines represent Eqs.~\eqref{eq03} and \eqref{eq05}. The macroscopic approach appears stationary while the conditional measurements reveal aged spectra.  }
		\label{fig:SuperpositionRTNmacro}
\end{figure}

In Fig.~\ref{fig:SuperpositionRTNmacro} we present the simulation results (symbols), and analytic results (solid lines) for the  two-state telegraph noise processes. The characteristic time scale $\tau$ varies from one realization to another with follows the PDF in Eq.~\eqref{PDF} with $\beta=1/2$, $\tau_{\rm min}=1$, and $\tau_{\rm max}=10^8$. We show the aging effect for the single-particle spectrum: the spectrum is reduced as we increase the measurement time, and the whole spectrum is shifted to the red since the lowest measured frequency is of the order of $1/t$ in agreement with  Eq.~\eqref{eq03}. Furthermore, finite-time simulation results show that the macroscopic approach appears stationary following Eq.~\eqref{eq05}.

It is rewarding that the superposition model, which is  probably the best well known model of $1/f^{\alpha}$ noise, shows aging if analyzed carefully. In contrast when we measure the macroscopic power spectrum, an apparently stationary spectrum is found. This resolve the conflict between  many empirical results, which found time-independent $1/f^{\alpha}$ spectrum (e.g. \cite{Hooge81,Dutta,Weissman}), and the nonstationary nature of $1/f^{\alpha}$ noise.  
Here, even though the macroscopic measured spectrum seems stationary, it still poses a finite power, see further discussion in Sec.~\ref{discussion}. 

We note that the spectrum which is measured in the single particle level, i.e Eq.~\eqref{eq03}, depends neither on $\tau_{\rm min}$ nor $\tau_{\rm max}$. The reason is simple; the PDF of the relaxation times, $P(\tau)\sim \tau^{-\beta}$  where $0<\beta<1$, must have an upper bound for convergence, while a lower cutoff is not necessary and can go to zero. The measurement time $t$ effectively serves as the upper cutoff. It means that the nature of the distribution, a heavy-tailed PDF, causes the measurement-time dependence of Eq.~\eqref{eq03} while the spectrum is independent of the inherent cutoffs, $\tau_{\rm min}$ and $\tau_{\rm max}$,  of the relaxation time distribution.   In App.~\ref{App.A} we discuss the cases where the $P(\tau)$ tail is ``less heavy'' in the sense that  $P(\tau)\tau$ decays to a constant or to zero for large $\tau$, i.e. $P(\tau)={\mathcal O}(1/\tau)$.

\subsection{Condition of $K$ Transitions}
\label{conditional}
\begin{figure}
	\centering
		\includegraphics[width=\columnwidth]{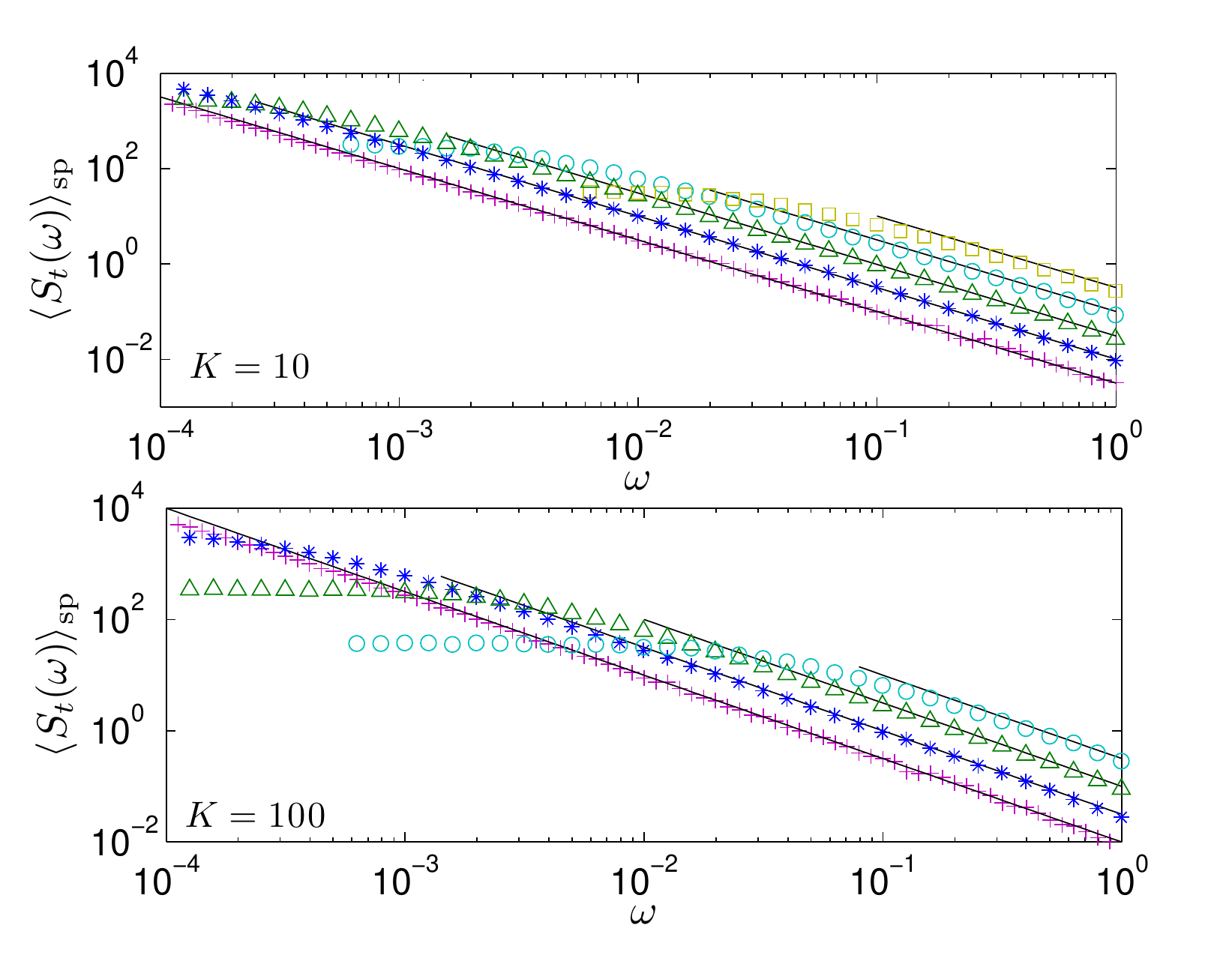}
		\caption{Single particle conditional spectrum, using the same parameters as in Fig.~\ref{fig:SuperpositionRTNmacro}, however now altering the condition on the number of transitions,
$K=10$ (upper panel) and $K=100$ (lower panel).  The aging effect is recovered however now we have a cutoff frequency $\omega_c\sim K/t$ below which we observe a flattening effect of the spectrum. Solid lines represent Eq.~\eqref{eq:22}.   }
		\label{RTNAgingK10}
\end{figure}

So far we defined the single molecule conditional measurement, based on the criterion of whether $I(t)$ jumped at least once from one level to another within the measurement time window  $[0,t]$. This conditional measurement is not unique,  and experimentally one may define other criteria, see further discussion in App.~\ref{OUclassification}.
However the main effect, an aging spectrum, is  generally valid. 
For example we define that a process is measured if the number of jumps between the two states is more than $K$ transitions, while a realization is not measured if the number of its transitions is less or equal to $K$. In particular the case $K=0$ was considered in the previous section.  
%
%
Then the probability of a realization with a characteristic sojourn time $\tau$ to be conditionally measured is
\begin{equation}
{\rm P}_{K}^{\rm mov}(t|\tau)=1-e^{-t/\tau}\sum_{k=0}^K \frac{[t/\tau]^k}{k!}=1-\frac{\Gamma(1+K,t/\tau)}{K!},
\label{conditionK}
\end{equation} 
and 
\begin{equation}
{\cal N}_t^{-1}\approx \frac{ \Gamma(\beta+K)\left(t/\tau_{\rm max}\right) ^{1-\beta}}{K!}
\end{equation} where the limit $\tau_{\rm min}\ll t \ll \tau_{\rm max}$ is taken. 
Here we obtain the power spectrum for microscopic measurements;
\begin{eqnarray}
\langle S_{t}(\omega) \rangle _{\rm sp}&& = {\cal N}_t I_0^2\int_{\tau_{\rm min}}^{\tau_{\rm max}}\frac{4\tau}{4+\omega^2\tau^2}
\label{eq:SP.K}
\\ \nonumber
&& \left(1-\frac{\Gamma(1+K,t/\tau)}{K!}\right)\frac{(1-\beta)\tau^{-\beta}}{\tau_{\rm max}^{1-\beta}-\tau_{\rm min}^{1-\beta}}{\rm d}\tau,
\end{eqnarray}
and recover the aging spectrum \eqref{eq03} with $A_\beta=K! 2^{1-\beta}(1-\beta)\pi\csc\left(\frac{\pi a}{2}\right)/\Gamma(\beta+K)$ for $0<\beta<1$, i.e. 
\begin{eqnarray}
&&\langle S_{t}(\omega)\rangle _{\rm sp}  \approx \label{eq:22}
\\ && I_0^2 2^{1-\beta}(1-\beta)\pi\csc\left(\frac{\pi a}{2}\right) \frac{K!}{\Gamma(\beta+K)} t^{\beta-1}\omega^{\beta-2} \nonumber
\end{eqnarray}
see Fig.~\ref{RTNAgingK10}. Detailed derivation is given in App.~\ref{App.k}. 
As mentioned, here we take into consideration only units with more than $K$ transitions, where $K\geq 1$. Those realizations are  effectively units with $\tau_j$ shorter than $t/K$, hence the spectrum has a natural cutoff at $\omega_c\sim K/t$. This means that the spectrum flattens when $\omega <\omega_c$. This effect is unique to the conditional spectrum and is not found for the macroscopic measurement since the latter is not sensitive to the measurement condition and it follows Eq.~\eqref{eq05} as before.      
The relation between the macroscopic spectrum and the conditional spectrum, Eq.~\eqref{eq04} holds for frequencies higher than the crossover frequency $\omega_c$ (see also App.~\ref{App.k}). 

\section{Ornstein-Uhlenbeck process} 
\label{OU}
Our observation of the aging effect in the single particle approach with conditional measurements is not limited to the two-state model.  
In the telegraph process in Sec.~\ref{RTN} we defined two populations via the number of transitions between the states. In a real data set the population can split into other categories, and in some cases the distinction between subsets of the populations is not obvious. 
For that reason 
we consider $N$ over-damped oscillators in contact with a thermal bath with temperature $T$. 
The process $I_j(t)$ is the position of the particle $j$,  which is modeled with the Ornstein-Uhlenbeck process $\dot{I_j} = -(m \omega_0^2/\gamma_j) I_j + \eta(t)$  \cite{uhlenbeck}. $\eta(t)$ is a white Gaussian noise with $\left\langle \eta(t)\eta(t')\right\rangle=2D\delta(t-t')$ where $D = k_BT/\gamma$ satisfies fluctuation-dissipation relation. The autocorrelation function of the $j$-th particle is 
\begin{equation}
\langle  I_j(t_0+t')I_j(t') \rangle = \frac{k_BT}{m\omega_0^2} \exp(-t'/\tau_j)
\end{equation}
with a relaxation time $\tau_j= \gamma_j/(m\omega_0^2)$ which is drawn from the mentioned PDF $P(\tau) \propto \tau^{-\beta}$ with $0<\beta<1$. 
For a particle $j$,  when $t\gg \tau_j$, the spectrum of the process is Lorentzian since then it is effectively stationary;
\begin{equation}
S_j(\omega)=\frac{k_BT}{m\omega_0^2}\frac{2\tau_j}{1+\omega^2\tau_j^2}.
\end{equation}
In the opposite limit, when $t\ll \tau_j$, this $j$-th spectrum is far from Lorentzian and appears random due to the nonergodic behavior on these time scales (see also App.~\ref{OUclassification}). 
Unlike the two-state process here we have two populations with distinct nontrivial spectra, i.e. we do not have noiseless units. 

\begin{figure}
	\centering
	\includegraphics[width=\columnwidth, trim= 0 140 0 00]{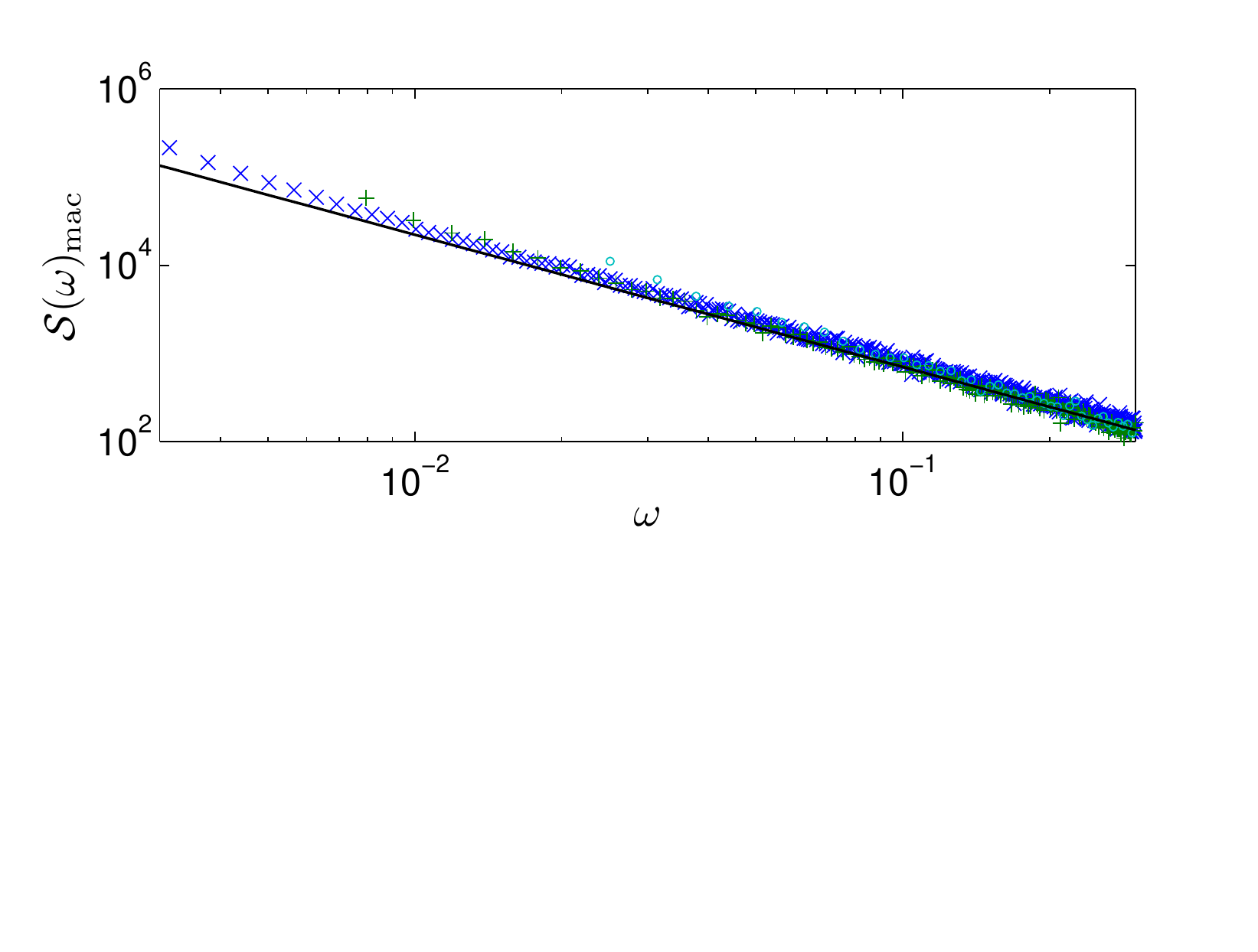}
		\includegraphics[width=\columnwidth, trim= 0 170 0 20]{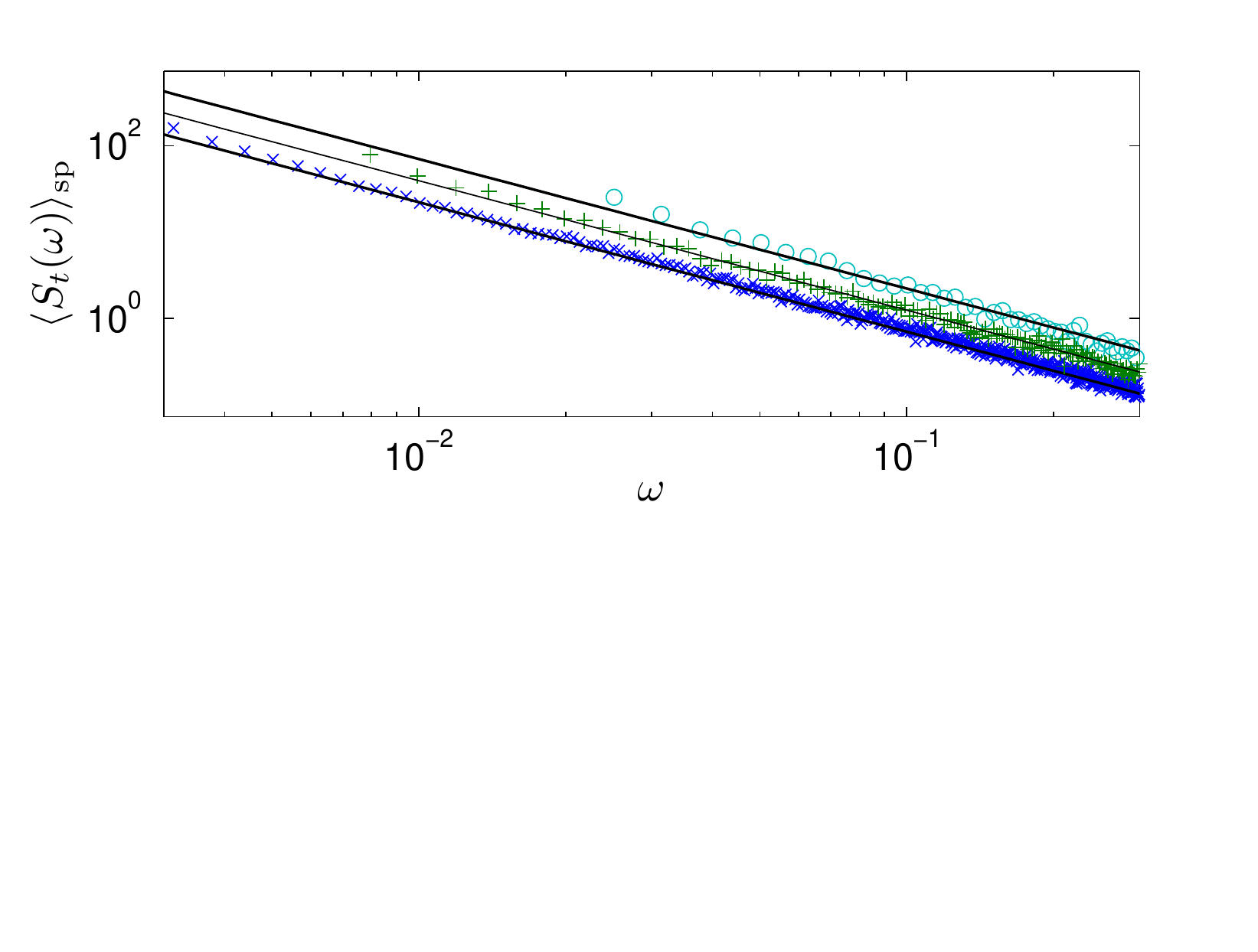}
		\caption{Simulation results for the Ornstein-Uhlenbeck process with $N=10^4$, $k_BT=1$ and $m\omega_0^2=1$. The relaxation times are fat-tailed distributed with $\beta=1/2$, $\tau_{{\rm min}}=1$ and $\tau_{{\max}} =10^6$. The measurement time of the spectrum was $t=10^3$ (cyan circles), $t=3162$ (green crosses) and $t=10^4$ (blue stars). The analytic predictions Eqs.~\eqref{eq:13} are presented in solid lines. The conditional measurements of spectra reveal an aging effect while the macroscopic approach obscures the nonstationarity.  }
		\label{fig:SuperpositionOUmacro}
\end{figure}

Here we distinguish between two populations; the first set contains realizations with $\tau_j<t$ which apparently exhibit Lorentzian spectra and in the second set the others with $\tau_j\geq t$ . The probability that a particle with a given relaxation time $\tau$ is measured
\begin{equation}
{\rm P}^{\rm mov}(t|\tau)=1-\Theta(t-\tau)
\end{equation}
where $\Theta(x)$ is the Heaviside function.   As in the previous model the number of particles with Lorentzian spectra increases with time, $N_t\approx N (t/\tau_{\rm max})^{1-\beta}$. Therefore the microscopic and the macroscopic measured spectrum present similar behavior as Eqs.~\eqref{eq03} and \eqref{eq05} (respectively)
with $A_\beta=B_\beta=(1-\beta)\pi\csc\left(\frac{\pi \beta}{2}\right)$  and $I_0^2=k_BT/(m\omega_0^2)$, namely we obtain
\begin{eqnarray}
\langle {S}(\omega)\rangle _{\rm sp}&\approx& \frac{k_BT}{m\omega_0^2} (1-\beta) \pi \csc\left(\frac{\pi\beta}{2}\right) t^{\beta-1}\omega^{\beta-2}, 
\label{eq:13} \\
{\cal S}(\omega)_{\rm mac}&\approx& N \frac{k_BT}{m\omega_0^2} (1-\beta) \pi \csc\left(\frac{\pi\beta}{2}\right) \tau_{\rm max}^{\beta-1}\omega^{\beta-2}.
\nonumber
\end{eqnarray}
In Fig.~\ref{fig:SuperpositionOUmacro} we present the simulation results where the power spectra of single particles (macroscopic samples) age (appear stationary). 

Optimization of single-molecule measurements and more advanced tools for distinguishing between populations is briefly discussed in App.~\ref{OUclassification}.

%
%

\section{Blinking quantum dot model}  
\label{BQD}
So far we have considered two models, where the underlying kinetics is stationary, in the sense that at least in principle, if we measure for an infinite time, the spectrum of each particle is Lorentzian.  Here we consider a stochastic model of blinking quantum dots. Those are nano-crystals that when interacting with a continuous wave laser field, emit light with intensity $I(t)$. The stream of photons emitted, blinks, and the process $I(t)$ exhibits on-off intermittency, with a power-law distribution of sojourn times in the on and off states. 
The power spectrum of single nano-crystals, measured {\em one at a time}, 
exhibits $1/f^{\alpha}$ fluctuations with clear nonstationary effects \cite{Sadegh,Margolin06,ferraro2009}. Here we focus on two unanswered questions: 
do we observe the aging effect in macroscopic measurements? and how do $S_{{\rm mac}}(\omega)$ and $\langle S_t(\omega)\rangle _{{\rm sp}}$ differ?

To answer these questions we define the underlining model. The signal, the light intensity, $I(t)$ takes two possible values, either $I(t)=I_0$ (state ``on'') or $I(t)=0$ (state ``off''). The blinking ``on'' $\leftrightarrow$ ``off''  sequence, for one dot,  is described by the set of ``on'' and ``off'' waiting times  $\left( {\cal T}_1^{{\rm on}}, {\cal T}_2 ^{{\rm off}}, ... \right)$. These sojourn times are statistically independent, identically distributed random variables with a 
common PDF  $\psi(\tau) \propto \tau^{-(1+\beta)}$. 
This model is a variant of both the trap model for dynamics in glasses  \cite{Bouchaud,bertin2002,bouchaud1992weak}  and of the velocity in L\'evy walk model  \cite{zaburdaev2015levy}. In particular \cite{Bouchaud} showed a nonstationary effect of the power spectrum of models of glassy dynamics, where the spectrum depends on the the waiting time $t_w$ defined below. In the blinking  model all the processes $I_j(t)$ are statistically identical, unlike the superposition model (Secs.~\ref{RTN}-\ref{OU}), where each unit has its own time scale associated with it. In what follows we assume $0<\beta<1$, hence the average waiting time diverges. 

\begin{figure}
	\centering
	\includegraphics[width=\columnwidth, trim= 0 140 0 0]{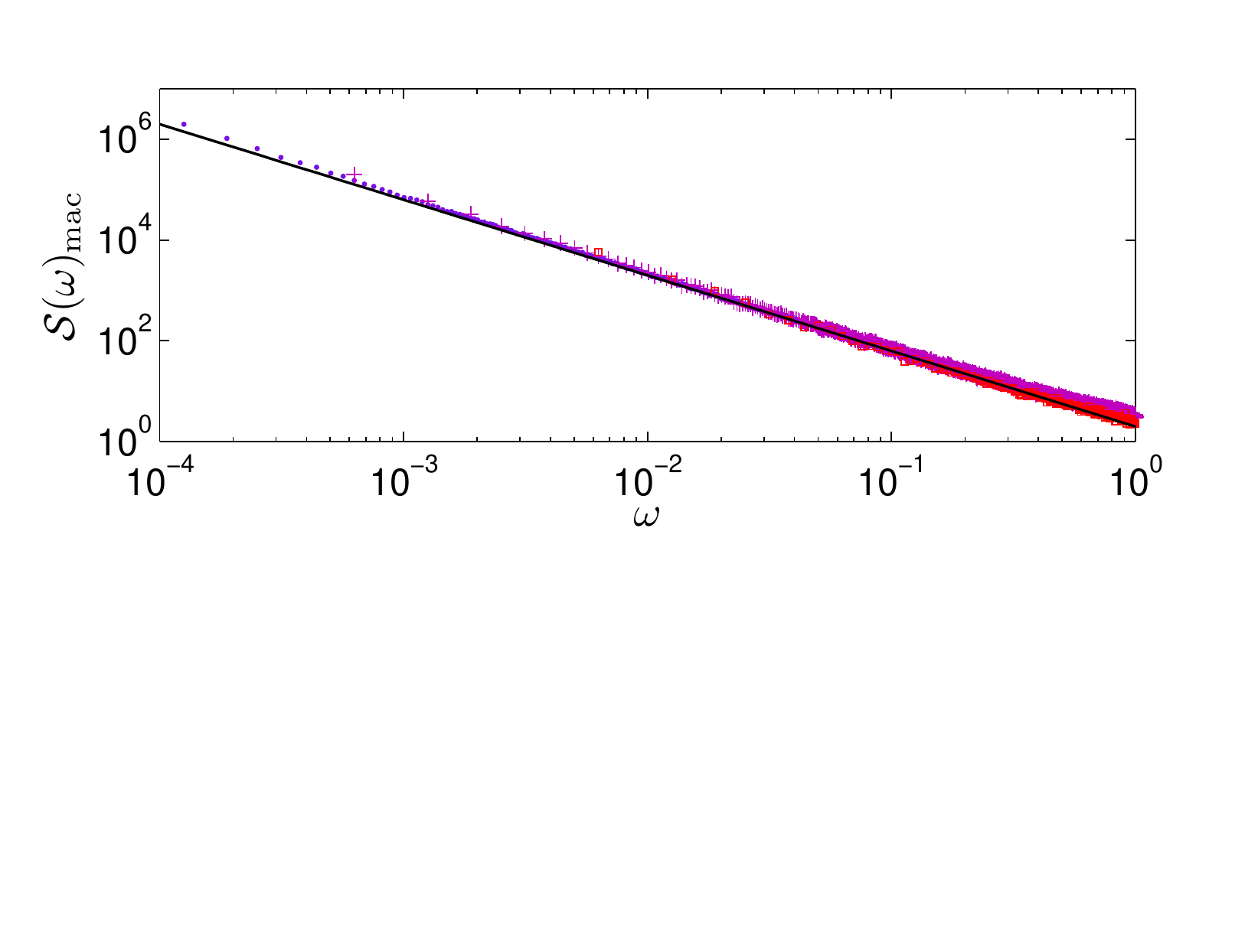}
		\includegraphics[width=\columnwidth, trim= 0 195 0 0]{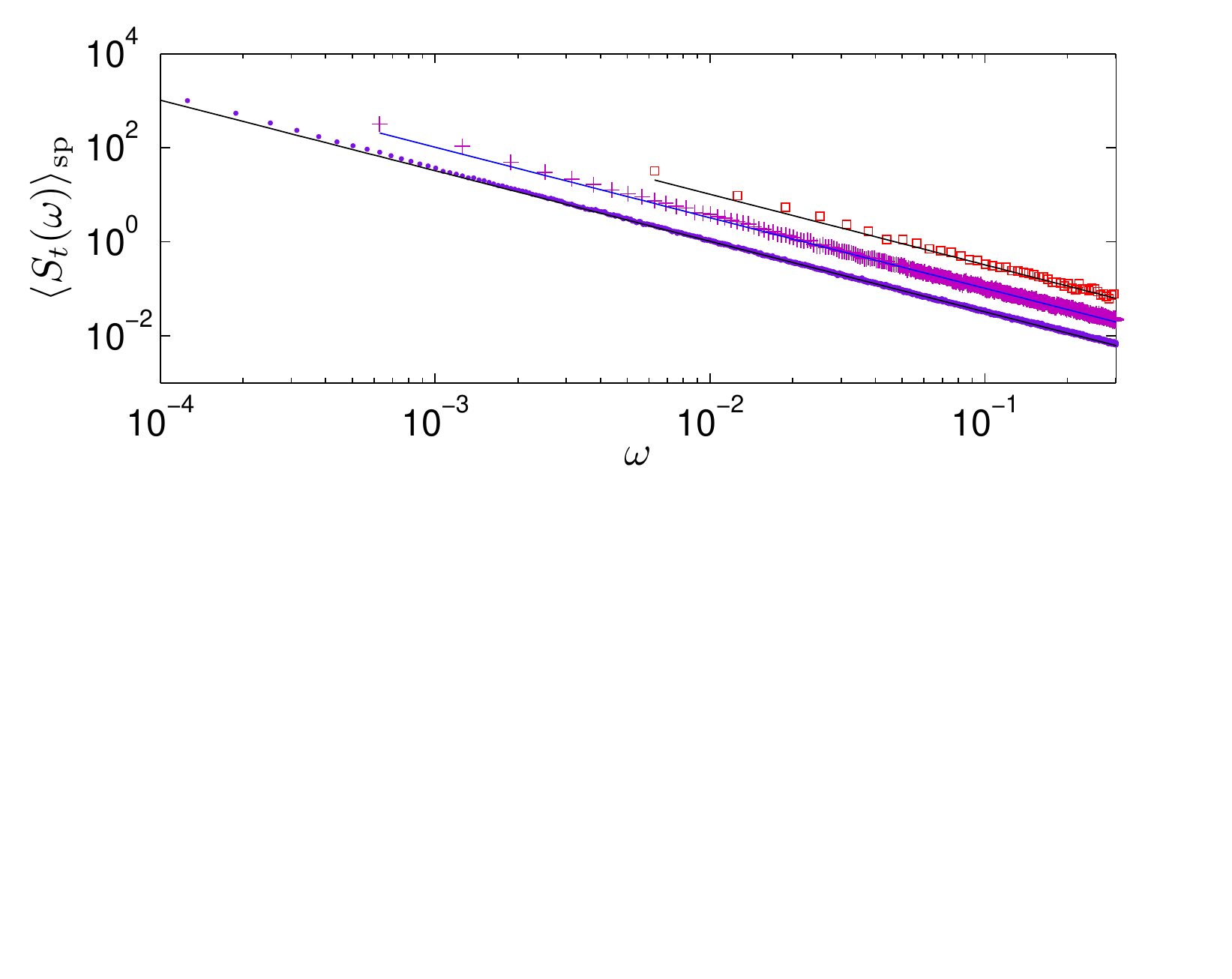}
		\caption{Simulation results for the blinking-quantum-dot model with $\beta=1/2$, $N=10^4$, $I_0=1$, $t_w=10^6$ at measurement time $t=10^3$ (red squares), $t=10^4$ (pink crosses) and $t=10^5$ (purple dots). The lines represent Eqs.~\eqref{eq06} and \eqref{eq08}. The conditional spectrum ages unlike the macroscopic measurements that appear stationary.  }
		\label{fig:SuperpositionBQDmacro}
\end{figure}

The $N$ processes are initially, at time $t=0$, in state ``on''. We wait a long time $t_w$ in which many transitions from ``on'' to ``off'' and vise versa take place. We then measure the spectrum by following the process in the time window $[t_w,t_w+ t]$, so $t$ is the measurement time. Also here we get two populations, a fraction of processes are jumping between the two states 
 in the time window of observation (the movers), while other processes are stuck. In the single realization level the idler's spectrum is zero. The movers are recorded in single molecule experiments 
and the conditional spectrum when $t\ll t_w$ reads \cite{Niemann2016}
\begin{equation}
\langle S_t (\omega) \rangle_{{\rm sp}} \simeq 
\frac{I_0^2}{2} \Gamma(2 - \beta) \cos\left( {\beta \pi \over 2} \right) t^{\beta-1} \omega^{\beta-2}. 
\label{eq06}
\end{equation}
The spectrum ages with the measurement time $t$ and is independent of the much longer waiting time $t_w$. To analyze the macroscopic measurement, we use a known formula for the probability to make at least one move in the time interval $[t_w, t_w + t]$ \cite{Godreche}, thus the average number of movers in the measured interval is
\begin{equation}
N_t  \simeq N \frac{\sin \pi \beta}{\pi (1 - \beta)} \left( \frac{t}{t_w} \right)^{1-\beta} 
\label{eq07}
\end{equation}
when $t/t_w \ll 1$. Here as we increase $t_w$, leaving $t$ fixed,  we get less and less moving processes. This is expected since the longer $t_w$ is, more and more processes get localized in one state in the observation window \cite{schulz2014}. 
Using Eq.~\eqref{eq04} we get a macroscopic spectrum which is measurement time independent \begin{equation}
{\cal  S}(\omega) _{\rm mac} \sim I_0^2 N \frac{\beta \cos \left(\frac{\pi \beta}{2} \right)}{
2 \Gamma(1 + \beta) } (t_w)^{\beta-1} \omega^{-2 + \beta}.
\label{eq08}
\end{equation}
Essentially this is similar to the superposition model, when we replace $\tau_{{\rm max}}$ with $t_w$, however the latter is a control parameter, together with the finite measurement time, in the experimental protocol.  In Fig.~\ref{fig:SuperpositionBQDmacro} we show the simulation (symbols) and analytic (lines) results of both single particle spectra and the macroscopic ones, where the distinction is made visual. 

To conclude in Eq.~\eqref{eq08} we see an aging spectrum in the spirit of the result of \cite{Bouchaud} in the sense that the spectrum depends on $t_w$. 
Eq.~\eqref{eq06} describes single particle measurements of the spectrum, and of course there is no contradiction between the two. 

\section{Intermittent map}
\label{Map}
The essential difference between the  method of measurements, the macroscopic and the single particle spectra, is not limited to stochastic processes. We have further extended the analysis to deterministic models, generating an intermittent signal from a Pomeau-Manneville type of map \cite{Manneville, pomeau1980}, and obtaining the spectra.

 We consider a deterministic signal generated by the following map; 
\begin{equation}
I_{t+1}={\cal M}(I_{t})
\end{equation}
where $t$ is a discrete time with unit time steps and the map is given by
\begin{equation}
{\cal M}(I_{t})=\left\{
\begin{array}{lll}
I_t+(aI_t)^{1+1/\beta} & 0\leq I_t<\xi_1 \\ \\
\frac{I_t-\xi_1}{\xi_2-\xi_1} & \xi_1\leq I_t\leq \xi_2  \\ \\
I_t-(a(1-I_t))^{1+1/\beta} & \xi_2<I_t\leq 1
\end{array}\right.
\label{eq:InteMap}
\end{equation} 
The signal is bounded, $0<I_t<1$ and $I_t$ is a function of a discrete time $t$. 
This map has two unstable fixed points, at $I_t=0$ and $I_t=1$. The discontinuities $\xi_1$ and $\xi_2$ are determined by $\xi_1+(a\xi_1)^{1+1/\beta}=1$ and $\xi_2+(a(1-\xi_2))^{1+1/\beta}=0$. The initial condition is uniformly distributed, and the process evolves via Eq.~\eqref{eq:InteMap}. Then we find for each realization its power spectrum corresponding to the signal recorded in the interval $[t_w,t_w+t]$.   

\begin{figure}
	\centering
		\includegraphics[width=1.05\columnwidth]{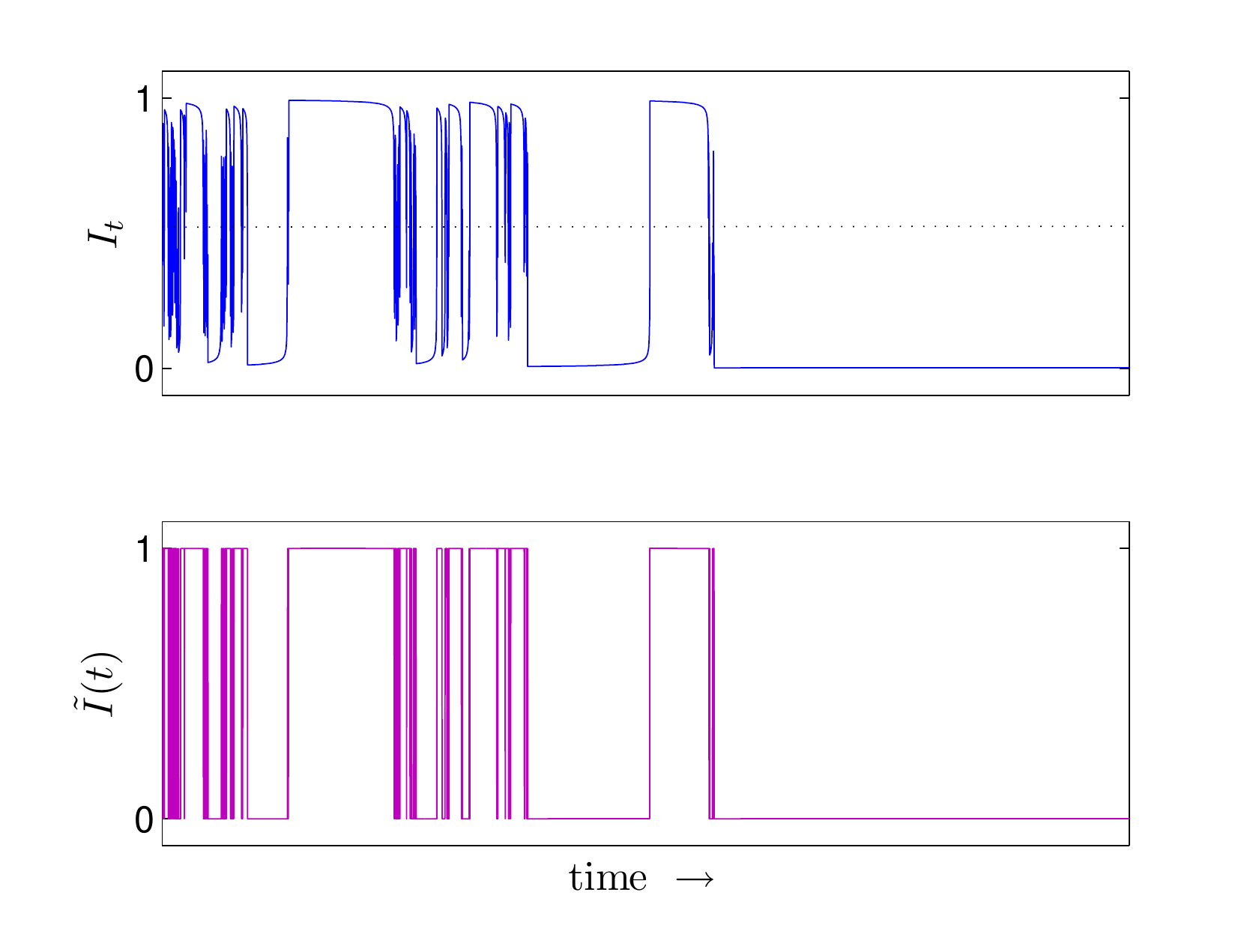}
		\caption{
			The deterministic signal $I_t$ (upper panel, blue) generated from the Pomeau-Manneville map  \eqref{eq:InteMap} with $a=1$ and $\beta=1/2$. A realization is considered as a mover where $I_t$ crosses the threshold $I^*=\frac{1}{2}$ (represented in a dashed line) at least once in the measurement-time period. The signal $I_t$ is modeled with a two-state stochastic process $\tilde{I}(t)$ (lower panel, pink).  }
	\label{fig:MapModel}
\end{figure}

\begin{figure}
	\centering
		\includegraphics[width=1.05\columnwidth]{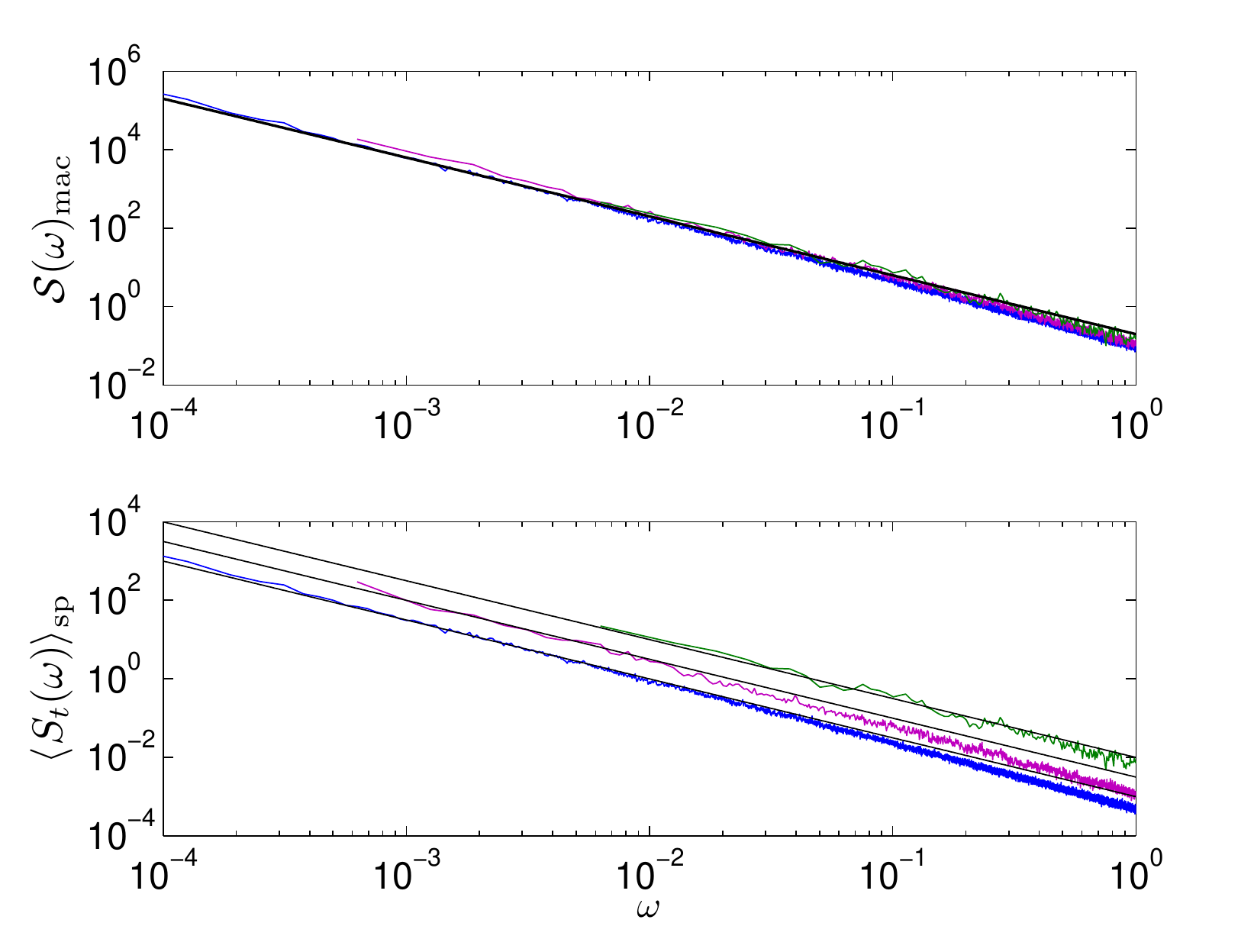}
		\caption{Comparison between macroscopic (top) and single particle measured spectrum (bottom panel) in the two-unstable-fixed points deterministic map \eqref{eq:InteMap}. The waiting time is $t_w=10^6$, $\beta=0.5$ and $N=10^3$, for three measurement times; $t=10^3$ (green), $t=10^4$ (pink) and $t=10^5$ (blue). The black lines represents Eqs.~\eqref{eq07} and \eqref{eq08}.     }
	\label{fig:MacroVsMicroMap}
\end{figure}

The signal $I_t$ exhibits a noisy on-off intermittency, due to the unstable fixed points. It is known \cite{Geisel,zumofen1993} that the PDF of the sojourn times in vicinity of each of the unstable fixed points is $\psi(\tau)\sim \tau^{-1-\beta}$ and that renewal theory discussed in Sec.~\ref{BQD} describes many properties of this deterministic process \cite{Geisel,zumofen1993}.
We distinguish between movers and idlers by the rule that if a signal crosses the threshold, e.g. $I^*=1/2$, at least once in the time interval $[t_w,t_w+t]$ it is considered a mover  (see Fig.~\ref{fig:MapModel}).  
We record $N=10^3$ realizations for averaging over the initial condition. In Fig.~\ref{fig:MacroVsMicroMap} we present the simulation results of the spectrum corresponding to the deterministic signals $I_t$. The macroscopic spectrum appears nonstationary while the conditional spectrum presents aging. This deterministic map is different, of course, if compared with the idealized stochastic on-off process discussed in previous section, see Sec.~\ref{BQD}. Still the predictions of the simple stochastic two state model Eqs.~\eqref{eq06} and \eqref{eq08} presented in black lines in Fig.~\ref{fig:MacroVsMicroMap}  seem to capture the main effects of aging.

\section{discussion}
\label{discussion}
\subsection{Convergence of the Total Power}
The appearance of $1/f^{\alpha}$ noise with $\alpha \geq 1$ seems at first glance unphysical due to the divergences $\int_{1/t}^{\infty}\omega^{-\alpha}{\rm d}\omega~=~\infty$ when $t\rightarrow\infty$. For an ergodic or a bounded process the total power must be finite from the following reason: the total power when $t\rightarrow \infty$ is  
$
\int_{1/t}^{\infty}S(\omega){\rm d}\omega \propto  \int_{0}^{t}I(t')^2{\rm d}t'/t
$
from Parseval identity, therefore for a stationary ergodic process $ \int_{0}^{t}I(t')^2{\rm d}t'/t= \langle I^2 \rangle$ which is finite. Furthermore, for a bounded process, i.e. $|I(t)|\leq I_{\rm max}$, regardless of whether it is ergodic or not, $ \int_{0}^{t}I(t')^2{\rm d}t'/t\leq I_{\rm max}^2 $, which is finite as well. This contradiction between the finite total power and the $1/f^{\alpha}$ measurements, is sometimes called the ``infrared catastrophe'' or  the ``1/f paradox'', e.g. \cite{Mandelbrot1967,solo,Niemann}, as was mentioned in the introduction. 

The time-dependent $1/f^{\alpha}$ noise solves the contradiction between the vastly measured $1/f^{\alpha}$ noise, and the convergence of the total power \cite{LeibovichLong}. 
In this paper we have found that the macroscopic measured spectrum appears stationary, i.e. it is time independent.
For example consider the random telegraph model presented in Sec.~\ref{RTN}. This model is bounded, $I_{\rm max}=I_0$, and its macroscopic measured spectrum is time independent ${\cal S}(\omega)_{\rm mac}\propto N (I_0)^2\omega^{-2+\beta}\tau_{\rm max}^{-1+\beta}$ following Eq.~\eqref{eq05}. Nevertheless it still poses finite power since
\begin{eqnarray}
&&\int_{1/t}^{\infty}{\cal S}(\omega)_{\rm mac}{\rm d}\omega\propto  N(I_0)^2\int_{1/t}^{\infty}\omega^{-2+\beta}\tau_{\rm max}^{-1+\beta}{\rm d}\omega= \nonumber\\
&&= N(I_0)^2\left(\frac{t}{\tau_{\rm max}}\right)^{1-\beta}\leq N(I_0)^2
\end{eqnarray}
in the limit $t\ll \tau_{\rm max}$. In the opposite limit, $t \gg \tau_{\rm max}$, the spectrum bends at frequency of order of $\tau_{\rm max}^{-1}$ and there is no low frequency divergence anyway. A similar result, $\int_{1/t}^{\infty}{\rm d} \omega{\cal S}(\omega)_{\rm mac}\leq N(I_0)^2$, is given for the quantum-dot model (Sec.~\ref{BQD}) when we replace $\tau_{\rm max}$ with $t_w$, and $t\ll~ t_w$.

\subsection{Superposition model}
\label{Superposition}
As was mentioned, a widely used model which generates $1/f^{\alpha}$ noise, originally suggested in the late 30's by Bernamont in the context of resistance fluctuations in thin films, 
is based on the superposition of many Lorentzian spectra \cite{bernamont,McWhorter,Dutta,Hooge81}. This is also called the distributed kinetics approach to $1/f^{\alpha}$ noise, and is probably the best well known explanation of the phenomenon.  As we consider in the Secs.~\ref{RTN} and \ref{OU}  the spectrum of unit $j$ is a measurement-time independent Lorentzian 
\begin{equation}
S_j(\omega) = \langle I^2 \rangle  \frac{ 2\tau_j}{ 1 + \omega^2 (\tau_j)^2 }
\label{eq01}
\end{equation}
 with the time scale $\tau_j$ varying from one unit $j$ to another with a common PDF 
 \begin{equation}
P(\tau) = {\cal N}  \tau^{-\beta}
\end{equation}
with $0<\beta<1$ and $\tau_{{\rm min}} < \tau <\tau_{{\rm max }}$. Then the normalization constant is ${\cal N} = (1-\beta) [ (\tau_{{\rm max}} )^{1-\beta} - (\tau_{{\rm min}})^{1- \beta}]^{-1}$.
By averaging over the spectrum Eq.~\eqref{eq01} we get an equation which serves as a starting point to many articles in the field \cite{Dutta,Hooge81,VanDerZiel}
\begin{equation}
\langle S(\omega) \rangle = {\cal N} \langle I^2 \rangle \int_{\tau_{{\rm min}}} ^{\tau_{{\rm max}} } \frac{2\tau}{1 + \omega^2 \tau^2} \tau^{-\beta} {\rm d} \tau.
\label{eq02}
\end{equation}
This formula must be used with care, since the spectrum  Eq.~\eqref{eq02} depends on $\tau_{{\rm max}}$, which is
unphysical in the context of $1/f^{\alpha}$ fluctuations,  for two reasons. 
It is clear that we must consider two cases, the first when the measurement time $t$ is shorter than $\tau_{{\rm max}}$. This is a typical situation, for example in glassy systems
 $\tau_{{\rm max}}$ was estimated to be of the order of the universe's age \cite{Crisanti,Bouchaud}. In this case Eq.~\eqref{eq02} does not describe neither macroscopic nor microscopic spectra,
 since it depends on the cutoff time $\tau_{\rm max}$ which is not detectable on the time scale of the experiment. 
The second option $t>\tau_{{\rm max}}$ is an experimental possibility, at least in principle, but if this holds we will {\em not} detect $1/f^{\alpha}$ noise at low frequencies \cite{diaz2015}. Namely
at frequencies of the order of $1/t$, one observes a flat spectrum in disagreement with the very basic definition of the phenomenon. Indeed many have searched for the bend down of $1/f^{\alpha}$ noise, mostly unsuccessfully, e.g. \cite{Caloyannides,Mandelbrot69}. 
 
 Therefore the power spectrum Eq.~\eqref{eq02} describes neither macroscopic nor microscopic measurements. 
On one hand Eq.~\eqref{eq02} does not depend on the ensemble size $N$, and cannot be considered as ``macroscopic''.  On the other hand Eq.~\eqref{eq02} does not represent a microscopic measurement since it is independent of the measurement time $t$, see Eqs.~\eqref{eq03},~\eqref{eq:22},~\eqref{eq:13}~and~\eqref{eq06}.

\subsection{A Note on Conditional Measurements}
 Today, with the advanced measurement techniques, one is able to record a signal from a single molecule or a nano-object, e.g. \cite{stefani2009beyond,metzler2014anomalous}.  
 Therefore the power spectrum of a microscopic unit, and further the conditional spectrum, becomes measurable. In  the context of measurement of diffusion of single molecules in the live cell, conditional measurements are routinely performed. In this situation one detects mixtures of spatially diffusing tracers and localized trapped particles, the diffusivity is conditionally measured, i.e. averaged with respect to the moving subpopulation \cite{schulz2014,Bronstein,Cherstvy}.  As we have showed here in the context of the $1/f^{\alpha}$ noise, these conditional measurements reveal aging effect in basic models of $1/f^{\alpha}$ noise

\section{Summary and Conclusions}
We have shown theoretically that an aging effect of the power spectrum is found in single-particle experiments, where measurements are conditional. However the aging is totally obscured by the ensemble averaging and it is not detected by macroscopic approaches. Our results are valid both for processes which exhibit stationarity at infinite time (e.g. random telegraph noise), and processes which are essentially nonstationary (e.g. quantum dot model). Certain aspects of the condition induce non-universal features, e.g. the cutoff $\omega_c=K/t$ (see Eq.~\eqref{eq:22}, while other features like an aging spectrum are robust and in that sense universal. In this sense our work is timely since today, with the advance of single molecule measurements,  the distinction between the two types of  measurements becomes important. Thus conditional measurements, with their peculiar distinction from macroscopic ensemble averages, must be considered as a separate class of measurement protocol.  This we hope solved one of the oldest conflicts in non-equilibrium statistical mechanics; the  nonstationary  scenario for $1/f^\alpha$ noise, which was clearly overlooked in many reviews in the field, is a valid description of even the most basic models of the field. 

\appendix
\renewcommand{\theequation}{{\thesection}\arabic{equation}} 

\section{Superimposed Lorentzian Spectra with other Relaxation-Time PDF $P(\tau)$}

\label{App.A}
\subsubsection{Derivation for $\beta=1$}
 $\beta=1$ is an important special case, since it gives exactly $1/f$ noise.
When $\beta=1$ we find the normalization constant ${\cal N}=\left[\ln(\tau_{\rm max})-\ln(\tau_{\rm min})\right]^{-1}$. Then the fraction of moving realizations is 
\begin{eqnarray} 
{\cal N}_t^{-1}&=&\int_{\tau_{\rm min}}^{\tau_{\rm max}}\left(1-e^{-t/\tau}\right)\frac{\tau^{-1}}{\ln(\tau_{\rm max})-\ln(\tau_{\rm min})}{\rm d}\tau  \nonumber \\ &\approx& \frac{\ln(t/\tau_{\rm min})}{\ln(\tau_{\rm max}/\tau_{\rm min})},
\end{eqnarray}
when $\tau_{\rm min}\ll t\ll \tau_{\rm max}$.
The single particle spectrum, hence, is given by
\begin{eqnarray}
\langle {S}(\omega)\rangle  _{\rm sp}&=&I_0^2\int_{\tau_{\rm min}}^{\tau_{\rm max}}\frac{4\tau}{4+\omega^2\tau^2}\left(1-e^{-t/\tau}\right)\frac{\tau^{-1}}{\ln(t/\tau_{\rm min})}{\rm d}\tau  \nonumber \\ 
&\approx& \frac{I_0^2\pi}{\ln(t/\tau_{\rm min})\omega},
\end{eqnarray}
and the macroscopic spectra reads
\begin{equation}
{\cal S}(\omega)_{\rm max}\approx \frac{N (I_0)^2\pi}{\ln(\tau_{\rm max}/\tau_{\rm min})\omega}.
\end{equation}
We conclude that when $\beta=1$, the spectrum depends on both bounds, upper and lower, of the relaxation times.

\subsubsection{Discussion about the case where $1<\beta<2$ and other relaxation-time distributions}
When $0<\beta<1$ we find that the low relaxation-time cutoff, $\tau_{\rm min}$ does not affect the asymptotic results. This, however, would not be the case where $1<\beta<2$. We assume fat-tailed relaxation time distribution with $P(\tau) \approx \left[(\beta-1)/\tau_{\rm min}^{1-\beta}\right]\tau^{-\beta}$
where the limit $\tau_{\rm min}\ll \tau \ll \tau_{\rm max}$ is taken.  The probability  that a realization with relaxation time $\tau$ moves in the measurement interval $[0,t]$ is ${\rm P}_0^{\rm mov}(t|\tau)=1-\exp(-t/\tau)$. Then the normalized distribution of the measured $\tau$ is $P(\tau){\rm P}_0^{\rm mov}(t|\tau){\cal N}_t$ where 
\begin{eqnarray}
{\cal N}_t^{-1}&=&\int_{\tau_{\rm min}}^{\tau_{\rm max}}{\rm d}\tau {\rm P}_0^{\rm mov}(t|\tau) P(\tau) =  \\ \nonumber
&\approx & \int_{\tau_{\rm min}}^{\tau_{\rm max}}{\rm d} \tau \left(1-e^{-t/\tau}\right) \frac{\beta-1}{\tau_{\rm min}^{1-\beta}}\tau^{-\beta}  \stackrel{ \tau_{\rm min}\ll t}{\longrightarrow}1
\end{eqnarray}
That means that the fraction of the measured particles converges to $1$, namely all particles are measured. Therefore we obtain 
\begin{equation}
\langle S(\omega)\rangle _{\rm sp}\approx I_0^2 2^{1-\beta}\pi\csc\left(\frac{\pi\beta}{2}\right)\tau_{\rm min}^{\beta-1}\omega^{\beta-2}
\end{equation}
which appears stationary.
 The intuitive explanation is the following. The relaxation times PDF $P(\tau)\propto \tau^{-\beta}$ rapidly decays at long $\tau$, and the fraction of units with long relaxation times is almost zero. Then the contribution to the spectrum from realizations with long relaxation times does not affect the spectra, and finite measurement time won't change the measured spectra. 

A stationary conditional spectrum is also found when the relaxation time distribution decays faster than $1/\tau$, i.e. when $P(\tau)=o(1/\tau)$ which means that $\lim_{\tau\rightarrow t^-} P(\tau)\tau\rightarrow 0 $. For example $P(\tau)$ follows Gaussian distribution or decays exponentially.

\section{Condition of $K$ Transitions}
\label{App.k}
We consider a blinking process which is defined by a two-state signal switching between $I(t)=+I_0$ and $I(t)=-I_0$. The sojourn times in each state are independent identically exponentially distributed random variables with characteristic mean $\tau_j$ for the $j$-th particle. Thus, for a given unit, we draw random waiting time ${\cal T}_1^j$  from the mentioned exponential distribution, the unit is in state $+I_0$ in the interval $[0,{\cal T}_1^j)$. Then we generate   ${\cal T}_2^j$ and then renew the process by switching to state $-I_0$ and so on. Then for a realization $j$ the process is defined by the array of random variables $\{{\cal T}_1^j,{\cal T}_2^j,{\cal T}_3^j,\ldots \}$. As mentioned in the text for unit $j$ the mean of the variables $\{{\cal T}^j\}$, $\tau_j$, is fixed, and varies from one unit to the other. 

The stationary correlation function of realization $j$ is 
\begin{equation}
\langle I_j(t)I_j(t+t') \rangle =I_0^2 \exp(-2t'/\tau_j)
\end{equation}
where the relaxation time is $\tau_j$. The corresponding spectrum is obtained from the Wiener-Khinchin theorem 
\begin{equation}
S_j(\omega)=I_0^2\frac{4 \tau_j}{4+\tau_j^2\omega^2}.
\label{S1}
\end{equation}
In our model the mean sojourn times $\{\tau_j\}$ are identical independent distributed random variables with probability density function 
\begin{equation}
P(\tau)={\cal N}\tau ^{-\beta}  \ \ \ \  \ \ \ \ \ \ \tau_{\rm min}<\tau<\tau_{\rm max}
\label{eq:PDF}
\end{equation}
with normalization constant ${\cal N}=(1-\beta)[\tau_{\rm max}^{1-\beta}-\tau_{\rm min}^{1-\beta}]^{-1}$ where $0<\beta<1$. 

The conditional microscopical measurement  includes only realizations that exhibit more than $K$ transitions in the time interval $[0,t]$. The probability of a given realization with mean waiting time $\tau$ to be measured is found using the Poisson distribution
\begin{equation}
{\rm P}_{K}^{\rm mov}(t|\tau)=1-\sum_{k=0}^K \frac{e^{-t/\tau}\left[t/\tau\right]^k}{k!},
\end{equation}
where in the case $K=0$ we find ${\rm P}_{0}^{\rm mov}(t|\tau)=1-~\exp[-t/\tau]$ as is given in Eq.~\eqref{eq:Condition}.
Then the normalization of the distribution of the active particles' relaxation times  reads
\begin{widetext}
\begin{eqnarray}
 {\cal N}_t^{-1}&=&\int_{\tau_{\rm min}}^{\tau_{\rm max}} \left(1-\sum_{k=0}^K \frac{e^{-t/\tau}(t/\tau)^k}{k!}\right)\frac{(1-\beta)}{\tau_{\rm max}^{1-\beta}-\tau_{\rm min}^{1-\beta}}\tau^{-\beta}{\rm d}\tau = \\  \nonumber
&=& 1-\sum_{k=0}^K \frac{(1-\beta)t^k}{k!(\tau_{\rm max}^{1-\beta}-\tau_{\rm min}^{1-\beta})}\int_{\tau_{\rm min}}^{\tau_{\rm max}} e^{-t/\tau}\tau^{-\beta-k}{\rm d}\tau = \\ \nonumber
&=& 1-\sum_{k=0}^K \frac{(1-\beta)t^{1-\beta}}{k!(\tau_{\rm max}^{1-\beta}-\tau_{\rm min}^{1-\beta})} \left[-\Gamma\left(\beta+k-1,\frac{t}{\tau_{\rm min}}\right)+\Gamma\left(\beta+k-1,\frac{t}{\tau_{\rm max}}\right)\right].
\end{eqnarray}
In the limit of $\tau_{\rm min}\ll t\ll \tau_{\rm max}$ we find
\begin{eqnarray}
{\cal N}_t^{-1}&\approx&1-\sum_{k=0}^K \frac{(1-\beta)t^{1-\beta}}{k!\tau_{\rm max}^{1-\beta}} \left[\Gamma\left(\beta+k-1\right)+\frac{t^{\beta+k-1}}{(1-\beta)\tau_{\rm max}^{\beta+k-1}}\right] = \\ \nonumber
&=& 1-(1-\beta)\left(\frac{t}{\tau_{\rm max}}\right)^{1-\beta}\sum_{k=0}^{K}\frac{\Gamma(\beta+k-1)}{k!}-\sum_{k=0}^K\frac{(t/\tau_{\rm max})^k}{k!} \\ \nonumber
&=& 1+\frac{\Gamma(K+\beta)}{K!}\left(\frac{t}{\tau_{\rm max}}\right)^{1-\beta}-\frac{e^{t/\tau_{\rm max}}\Gamma\left(1+K,\frac{t}{\tau_{\rm max}}\right)}{K!}
\end{eqnarray}
where we use the relation $\sum_{k=0}^Kx^k/k!=e^x \Gamma (K+1,x)/K!$ and $\Gamma(a,z)=\int_z^{\infty}t^{a-1}e^{-t}{\rm d}t$ is the incomplete Gamma function. Therefore, in the limit $t \ll \tau_{\rm max}$ we obtain
\begin{equation}
 {\cal N}_t^{-1}\approx\frac{\Gamma(K+\beta)}{K!}\left(\frac{t}{\tau_{\rm max}}\right)^{1-\beta}.
\end{equation}
When $K=0$ we recover ${\cal N}_t^{-1}\approx\Gamma(\beta)\left({t}/{\tau_{\rm max}}\right)^{1-\beta}$.
Following Eq.~\eqref{eq:SP.K} the conditional microscopic spectrum, thus, is 
\begin{equation}
\langle {S}_t(\omega)\rangle  _{\rm sp}
\approx
I_0^2\int_{\tau_{\rm min}}^{\tau_{\rm max}}\frac{4\tau}{4+\omega^2\tau^2}\left(1-\frac{\Gamma\left(1+K,\frac{t}{\tau}\right)}{K!}\right)\frac{(1-\beta)\tau^{-\beta}}{{\cal N}_t^{-1}(\tau_{\rm max}^{1-\beta}-\tau_{\rm min}^{1-\beta})}{\rm d}\tau, 
\end{equation}
Substitute ${\cal N}_t$ into that and expand the integration interval to $[0,\infty)$ where the limit 
$\tau_{\rm min}\ll t\ll \tau_{\rm max}$ is considered
\begin{eqnarray}
\langle S_t(\omega)\rangle_{\rm sp}&\approx& I_0^2\int_{0}^{\infty}\frac{4\tau}{4+\omega^2\tau^2}\left(1-\frac{\Gamma\left(1+K,\frac{t}{\tau}\right)}{K!}\right)\frac{(1-\beta)K!}{\Gamma(K+\beta) t^{1-\beta}}\tau^{-\beta}{\rm d}\tau.
\end{eqnarray}
Integrating using Mathematica gives
\begin{eqnarray}
\langle S_t(\omega)\rangle_{\rm sp}\approx && I_0^2\frac{2^{1-\beta}(1-\beta)t}{\Gamma(K+\beta)}\cdot \label{S9}\\ \nonumber 
&& \left\{\frac{ \Gamma (\beta+K-1) \,  _2F_3\left[1,1-\frac{\beta}{2};2-\frac{\beta}{2},-\frac{\beta}{2}-\frac{K}{2}+1,-\frac{\beta}{2}-\frac{K}{2}+\frac{3}{2};-\frac{1}{16} (\omega t) ^2\right]}{2-\beta} \right. \\ \nonumber
	&&-\frac{\pi  2^{-\beta-K} ( \omega t)^{\beta+K-1} \csc \left(\frac{1}{2} \pi  (\beta+K-1)\right) \,  _1F_2\left[\frac{K}{2}+\frac{1}{2};\frac{1}{2},\frac{K}{2}+\frac{3}{2};-\frac{1}{16} (\omega t)^2\right]}{K+1} \\ \nonumber
	&& \left.+\frac{\pi  2^{\beta-K-1} 
   (\omega t)^{\beta+K} \sec \left(\frac{1}{2} \pi  (\beta+K-1)\right) \, _1F_2\left[\frac{K}{2}+1;\frac{3}{2},\frac{K}{2}+2;-\frac{1}{16}
   (\omega t)^2\right]}{K+2} \right\}
\end{eqnarray} 
expanding for $\omega t\gg 1$ yields
\begin{equation}
\langle S_t(\omega)\rangle _{\rm sp}\approx I_0^2 2^{1-\beta}(1-\beta)\pi\csc\left(\frac{\pi\beta}{2}\right)\frac{K!}{\Gamma(K+\beta)}t^{\beta-1}\omega^{\beta-2}. \label{eq:SPK}
\end{equation}
This result is given in Eq.~\eqref{eq03} for $K=0$ with $A_{\beta}=2^{1-\beta}(1-\beta)\pi\csc\left(\pi\beta/2\right)/\Gamma(\beta)$. More generally for $K>0$ we find $A_{\beta}=2^{1-\beta}(1-\beta)\pi\csc\left(\pi\beta/2\right)\Gamma(K+1)/\Gamma(K+\beta)$. We use these results in Fig.~\ref{fig:SuperpositionRTNmacro} $(K=0)$ and Fig.~\ref{RTNAgingK10} $(K>0)$. 

Mathematically taking the opposite limit $\omega t \ll 1$ of Eq.~\eqref{S9} gives when $K\geq1$
\begin{equation}
\langle S(\omega t) \rangle _{\rm sp}\approx \frac{t}{(2-\beta)(K+\beta-1)},
\label{eq:S0}
\end{equation}
which is frequency  independent, hence, the spectrum bends from the $1/f^{\alpha}$ behavior, see Fig.~\ref{RTNAgingK10} in the main text. The crossover frequency $\omega_c$ is the frequency for which expression \eqref{eq:SPK} is equal to \eqref{eq:S0} and is given by
\begin{equation}
\omega_c \sim \frac{1}{t}\left(\frac{\Gamma(K+1)\pi\csc\left(\beta\pi/2\right)2^{1-\beta}(1-\beta)(2-\beta)}{\Gamma(K+\beta-1)}\right)^{-2+\beta} \stackrel{\longrightarrow}{_{K\gg 1}}  \frac{2K}{t} \left[\Gamma(\beta/2)\Gamma(2-\beta/2)(1-\beta)\right]^{-2+\beta}.
\end{equation} 
We conclude that the conditional spectrum reveals a flattening of the $1/f^{\alpha}$ behavior at frequencies lower than $\omega_c$, see discussion and Fig.~\ref{RTNAgingK10}.

The macroscopic spectrum does not depend on the measurement condition and is given by
\begin{equation}
{\cal S}(\omega)_{\rm mac}\approx N(I_0)^2 2^{1-\beta}(1-\beta)\pi\csc\left(\frac{\pi\beta}{2}\right)(\tau_{\rm max})^{\beta-1}\omega^{\beta-2}.
\end{equation}
The relation between the macroscopic spectrum to the conditional spectrum, ${\cal S}(\omega)_{\rm max}=N_t\langle S_{t}(\omega)\rangle_{\rm sp}$, holds for  frequencies higher than the crossover frequency $\omega_c$.

\section{Data Analysis in the Ornstein-Uhlenbeck Process }
\label{OUclassification}
In Sec.~\ref{OU} and in the simulation results presented in Fig.~\ref{fig:SuperpositionOUmacro} we use the following condition to separate between two populations:
is the relaxation time shorter than the measurement time or not. This method has the advantage that the number of particles in the measured set  is easily calculated. Then the microscopic spectra can be quantified and a comparison between simulation results and Eq.~\eqref{eq:13}
is presented in Fig.~\ref{fig:SuperpositionOUmacro}. However, in an experimental situation those relaxation times $\{\tau_j\}$ are {\em a priory} unknown. In the following we suggest two other methods which are more practical to use in an experimental scenario.

One criterion to distinguish between the populations is based on whether the variance of $I(t)$,
 on the time scale of the measurement $t$ is roughly given by the equipartition theorem. One may argue that particles which do not obey this rule have not reached  equilibrium until the measurement time $t$. This thermal criterion, which may serve as a benchmark  for conditioning the spectrum,  is not unique. 

A second procedure is based on the spectrum itself and hence more detailed.  
Each individual realization's spectrum $S_j(\omega)$ is fitted to a Lorentzian shape $g_L(\omega)$ and to a spectrum of a Brownian particle (this is reasonable since the particles with large $\tau$ are freely diffusing) $g_B(\omega)$ with a fitting parameter $\hat{\tau}_j$,
\begin{equation}
g_L(\omega)=\frac{2\tau_j}{1+\omega^2 (\hat{\tau}_j^L)^2} \ \ \ \ \ \ \ \ \ \ \ g_B(\omega)=(\hat{\tau}_j^B)^{-1}\omega^{-2}.
\end{equation} 

{\bf Confidence Interval:}
The first method for classification relies on the confidence interval. For each fitting model, $g_L$ and $g_B$, we get the fitting parameter $\hat{\tau}^{L}_j$ corresponds to model $g_L$ and $\hat{\tau}^{B}_j$ corresponds to $g_B$ and with a $95\%$ confidence in an interval $(a_{L},b_{L})$ and $(a_{B},b_{B})$ respectively,

\begin{table}[h]
	\centering
		\begin{tabular}{rccc} 
			General model: & $g_L(\omega)$ &\  & $g_B(\omega)$ \\
			Coefficients (with $95\%$ confidence bounds): &  $\hat{\tau}_j^L \ \ (a_L, b_L)$ &\ & $\hat{\tau}_j^B \  \ (a_B, b_B)$\\ 		
		\end{tabular}
\end{table}

We characterize the goodness of the Lorentzian fit by the width of the confidence interval ${\rm ci}_{L}=|b_{L}-a_{L}|/\hat{\tau}^{L}_j$, and similarly for the Brownian spectrum with ${\rm ci}_{B}=|b_{B}-a_{B}|/\hat{\tau}^{B}_j$. We classify a realization as Lorentzian when ${\rm ci}_L < {\rm ci}_B$ and as Brownian otherwise. 

We note that such a classification method needs to used with care since the regression hypothesis is not linear, and a non-convex cost function may appears. In that case, the fitted parameter $\hat{\tau}$ may be affected by the initial searching point. Here, a deeper analysis is needed, and we leave it for a future publication.
%
%
%
  
\begin{figure}
	\centering
		\includegraphics[width=0.60\columnwidth]{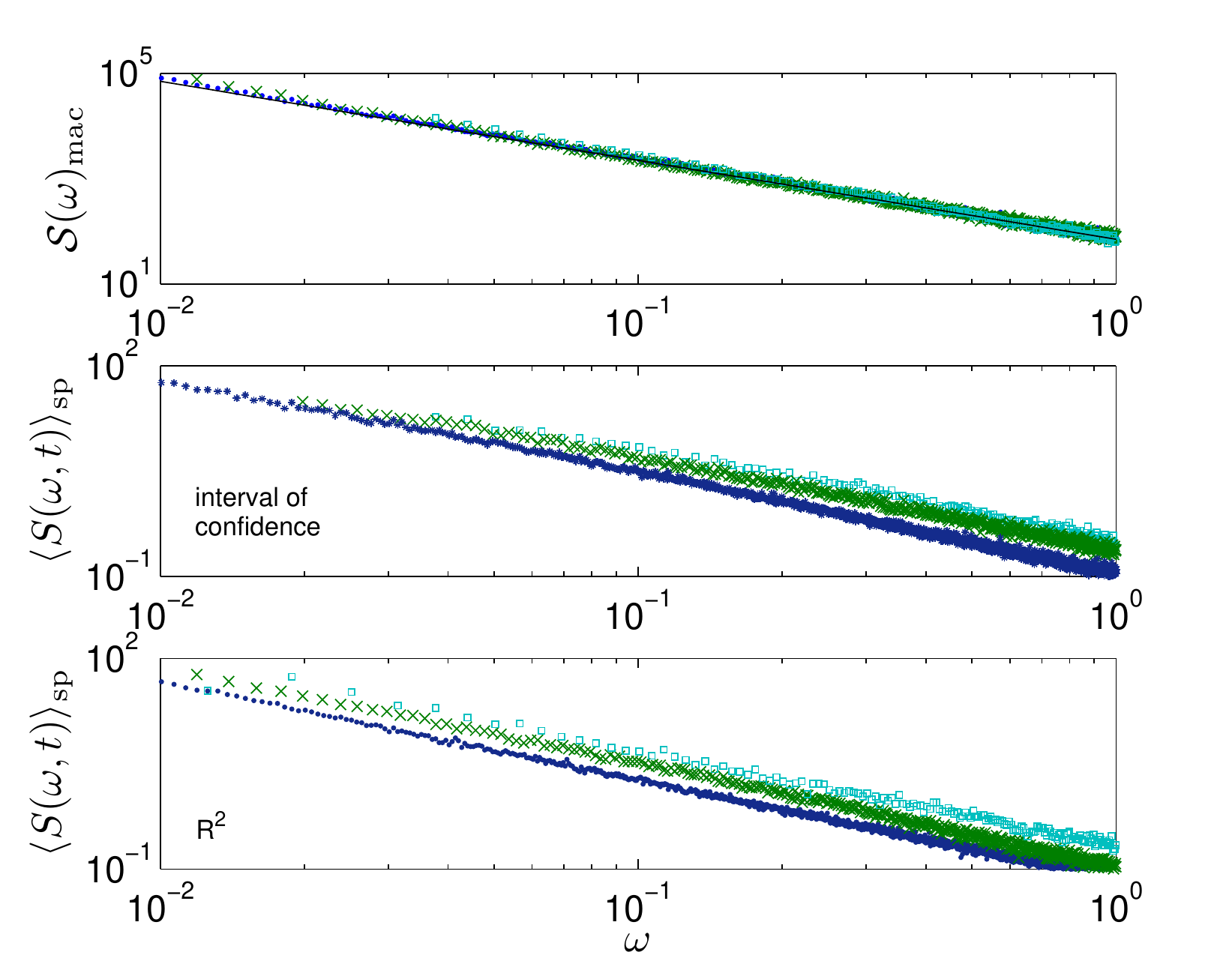}
		\caption{The simulation results for the Ornstein-Uhlenbeck Process. In the upper panel we present the macroscopic measured spectra at three measurement times; $t=10^3$, $3162$ and $10^4$. The black line represents the analytic prediction Eq.~\eqref{eq:13}. In the lower panel we present the microscopic measured spectra where we use the spectrum-based method for classification. The main result, the aging effect, is clearly visible in both classification methods.         }
	\label{fig:OU_spectra_classification}
\end{figure}

{\bf Coefficient of Determination:}
A second method to determined the goodness of the fitting model is related to the coefficient of determination, 
\begin{equation}
R^2=1-\frac{\sum_{i=1}^n \left(S_j(\omega_i)-g(\omega_i)\right)^2}{\sum_{i=1}^n \left(S_j(\omega_i)-\overline{S_j}\right)^2}
\end{equation}
where $\overline{S_j}=\sum_{i=1}^{n} S_j(\omega_i)/n$ and $n$ is the number of observed frequencies. $R^2\in [0,1]$ where $R^2$ is close to $1$ means reasonably a good fitting. For a realization $j$ we accepted the Lorentzian assumption where  $R_L^2>R_B^2$ and reject this assumption otherwise.

In Fig.~\ref{fig:OU_spectra_classification} we present the simulation results for the macroscopic and microscopic spectra with the two classification methods suggested above. The macroscopic spectra, of course, does not depend on the measurement criteria and left unchanged (upper panel). For the single particle spectrum, both criteria reveal aged spectrum (middle and lower panels) which is clearly visible in Fig.~\ref{fig:OU_spectra_classification}. The conclusion is that while conditional spectrum is a rather general concept which depends on the choice of the experimentalists, still the main conclusions in the text are robust.

\section{Blinking quantum dot model }
We consider a two-state process, i.e. ``on'' and ``off'' sequence,  where the waiting times at each state are fat-tailed distributed with PDF $P(\tau)\sim\tau^{-1-\beta}$. The process switches to the other state after a sojourn time. Unlike the two-state model with exponential waiting times, here all units are statistically identical. However none of them is stationary since $0<\beta<1$ implies the divergence of the mean sojourn time.
 
The macroscopic spectrum measured in a time interval $[t_w, t_w+t]$ for this two-state signal is \cite{Niemann2016}
\begin{equation}
{\cal S}(\omega)_{\rm mac} \approx N I_0^2 \frac{\cos(\pi\beta/2)}{2\Gamma(1+\beta)})\Lambda_{\beta}\left(\frac{t_w}{t}\right) t^{\beta-1}\omega^{\beta-2},
\end{equation}
where the aging factor is $\Lambda_{\beta}(x)=(1+x)^{\beta}-x^{\beta}$.
Therefore, in the limit $t_w\gg t$ we recover Eq.~\eqref{eq08} in the text
\begin{equation}
{\cal S}(\omega)_{\rm mac} \approx NI_0^2 \frac{\beta \cos(\pi\beta/2)}{2\Gamma(1+\beta)}t_w ^{\beta-1}\omega^{\beta-2}.
\label{eq:BQDmac}
\end{equation}
The probability of at least one transition in the measurement-time interval $[t_w,t_w+t]$ is \cite{Godreche}
\begin{equation}
P_{0}^{\rm mov}(t|t_w)=\frac{\sin(\pi\beta)}{\pi(1-\beta)}\left(\frac{t}{t_w}\right)^{1-\beta}{_2F_1}\left[1,1-\beta,2-\beta,-\frac{t}{t_w}\right] \stackrel{\longrightarrow}{_{t\ll t_w}} \frac{\sin(\pi\beta)}{\pi(1-\beta)}\left(\frac{t}{t_w}\right)^{1-\beta},
\end{equation}
and Eq.~\eqref{eq06} in the main text is recovered.  

\section{Paralleled Measurements}
\label{Paralleled}
We measured the spectrum for each realization; $S_{j}(\omega)$. These spectra are recorded simultaneously in parallel. Then the macroscopic spectrum is
\begin{equation}
{\cal S}(\omega)_{\rm mac}=\sum_{j=1}^N S_{j}(t,\omega)
\end{equation}
However, in some experimental cases, the macroscopic spectrum is measured via the macroscopic signal, 
\begin{equation}
\tilde{{\cal S}}(\omega)_{\rm mac}=\frac{1}{t}\left|\int_{0}^{t}I(t')e^{\imath\omega t'}{\rm d}t'\right|^2=
\frac{2}{t}\int_0^t \int_0^{t-t_1} I(t_1+\tau)I(t_1) \cos(\omega\tau) {\rm d}\tau {\rm d}t_1
\end{equation}
where the macroscopic signal is $I(t)=\sum_j I_j(t)$. Therefore
\begin{eqnarray}
\tilde{{\cal S}}(\omega)_{\rm mac}&=&
\frac{2}{t}\int_0^t \int_0^{t-t_1} \left(\sum_j I_j(t_1+\tau)\right)\left(\sum_{j'} I_{j'}(t_1) \right)\cos(\omega\tau) {\rm d}\tau {\rm d}t_1=\\ \nonumber
&=&\frac{2}{t}\int_0^t \int_0^{t-t_1} \sum_j I_j(t_1+\tau)I_{j}(t_1) \cos(\omega\tau) {\rm d}\tau {\rm d}t_1+\frac{2}{t}\int_0^t \int_0^{t-t_1} \sum_{j}\sum_{j'\neq j} I_j(t_1+\tau) I_{j'}(t_1)\cos(\omega\tau) {\rm d}\tau {\rm d}t_1 = \\ \nonumber
&=& \sum_{j=1}^N S_j(t,\omega) + \frac{2}{t}\int_0^t \int_0^{t-t_1} \sum_{j=1}^N\sum_{j'\neq j} I_j(t_1+\tau) I_{j'}(t_1)\cos(\omega\tau) {\rm d}\tau {\rm d}t_1.
\end{eqnarray}
Now, we claim that the second term on average is small, since the realizations are mutually independent with zero mean
\begin{equation}
\left\langle \frac{2}{t}\int_0^t \int_0^{t-t_1} \sum_{j}\sum_{j'\neq j} I_j(t_1+\tau) I_{j'}(t_1)\cos(\omega\tau) {\rm d}\tau {\rm d}t_1\right\rangle=0.
\end{equation}

In Fig.~\ref{fig:ParalleledMesured} we present the paralleled measured spectrum versus the spectrum correspond to the macroscopic signal in the random-telegraph-noise model and in the Ornstein-Uhlenbeck process. For the simulation we use $N=10^4$ particles, $P(\tau)\propto \tau^{-1/2}$ where $\tau\in[1,10^5]$ and the measurement time is $t=10^4$. The spectrum correspond to the total macroscopic signal $I(t)$ is somewhat noisy (represented in blue line), hence we smooth it with moving average with logarithmic-width windows (red stars). The summation of the paralleled measures spectra is represented in green line. We conclude that the paralleled measuring of the spectrum presents an agreement with the spectrum related to the total signal generated by the system.

\begin{figure}
	\centering
\includegraphics[width=0.48\columnwidth]{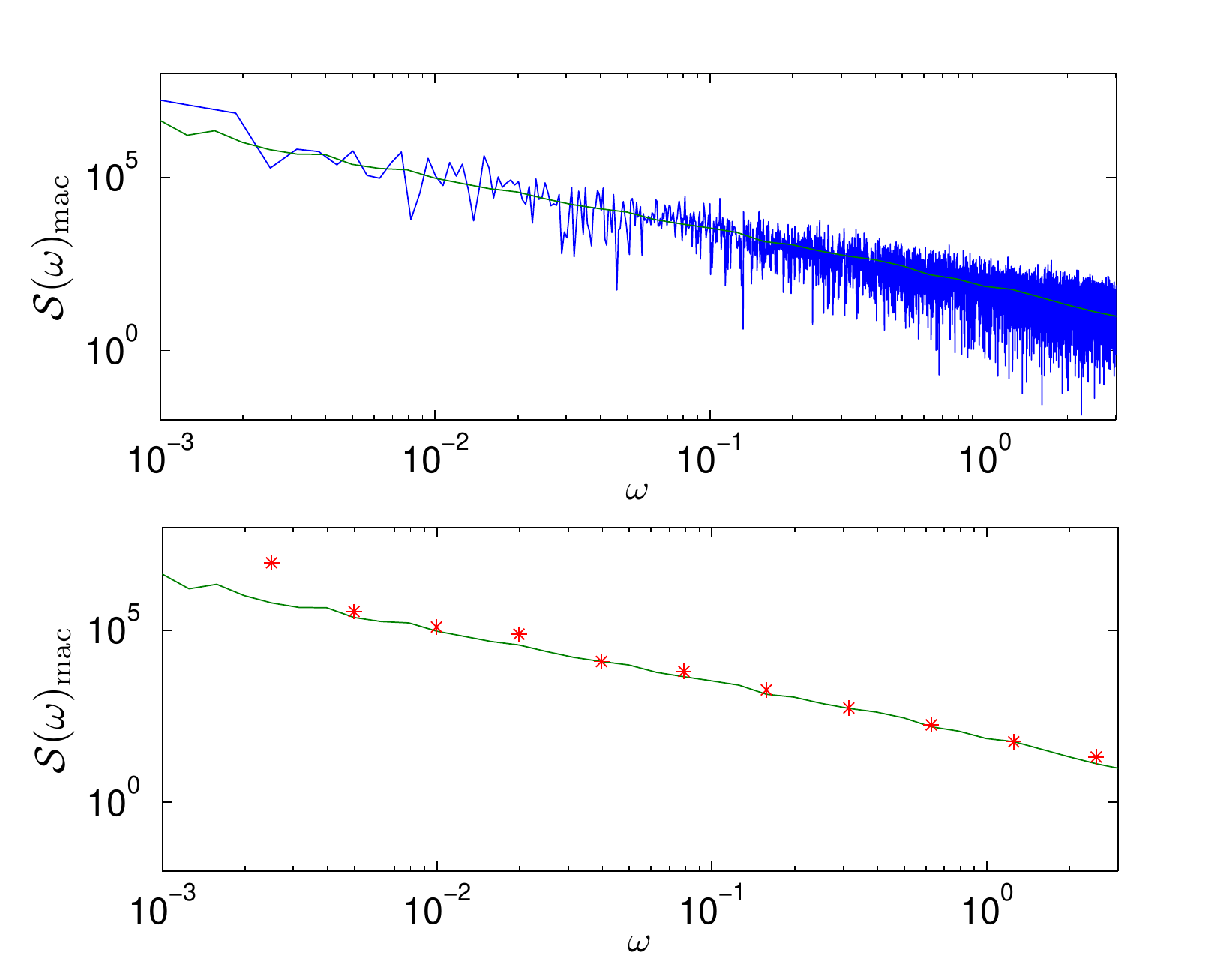}
\includegraphics[width=0.48\columnwidth]{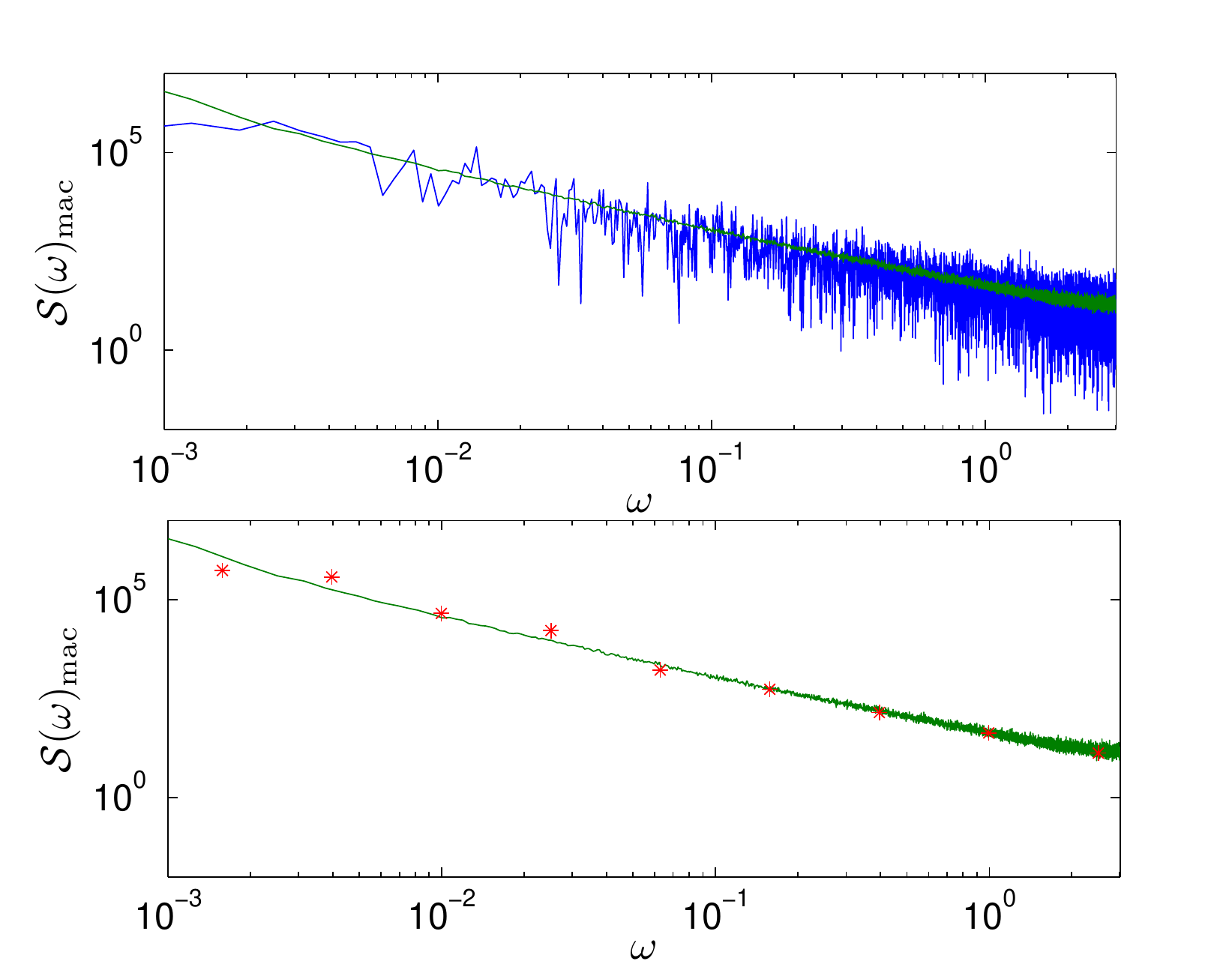}
		\caption{Upper panels: Paralleled measured power spectrum (green line) versus the total spectrum corresponds to the macroscopic signal (blue line) in two processes; the two-state random telegraph noise with $\beta=1/2$, $\tau_{\rm min}=1$, $\tau_{\rm max}=10^5$, $N=10^4$ and $t=10^4$ (left panels) and Ornstein-Uhlenbeck process with the same parameters (right) . Lower panels: The smoothed total spectrum (red stars) versus the paralleled measured spectrum (green line) shows that the parallel measurements provide a reasonable approximated measuring method. }
	\label{fig:ParalleledMesured}
\end{figure}

\end{widetext}

\section{Spectrum's Contribution from neglected particles}
\label{App.B}

\subsubsection{Two-State model}

A particle is considered trapped when its physical quantity $I(t)$ is a constant during the measurement period. In such a case the corresponding sample spectrum is
\begin{equation}
S_t(\omega)=\frac{I_0^2}{t}\left|\int_0^{t} \exp(-\imath\omega t'){\rm d}t' \right|^2=
 I_0^2\frac{4\sin ^2(\omega t/2)}{\omega^2 t}.
\end{equation}
$S_t(\omega)$ vanishes at the natural frequencies, $\omega_n=2\pi n/t$ for integer $n$. 

\subsubsection{Ornstein-Uhlenbeck Process}
In the two-state model, we consider a  particle as trapped when  $I(t)={\rm const.}$  over the measurement interval. For the Ornstein-Uhlenbech process one needs to determine whether a given particle is trapped or not. Generally, a trapped particle is consider when the friction force is very strong. In that case, when $\tau$ is very small, the process is stationary and the spectrum exhibits Lorentzian. In the other limit, for very weak friction force, i.e. long relaxation time, the particle is nearly diffusive, since it is not affected by the friction. The corresponding power spectrum is the one of the Brownian noise, i.e. $S_j(\omega) \sim \omega^{-2}$, with a prefactor which is proportional to the diffusion constant $D=k_BT/(m\omega_0\tau_j)$.

\subsubsection{Blinking Quantum Dot with Additional White Noise }

 \begin{figure}[t!]
 	\centering
 		\includegraphics[width=0.95\columnwidth]{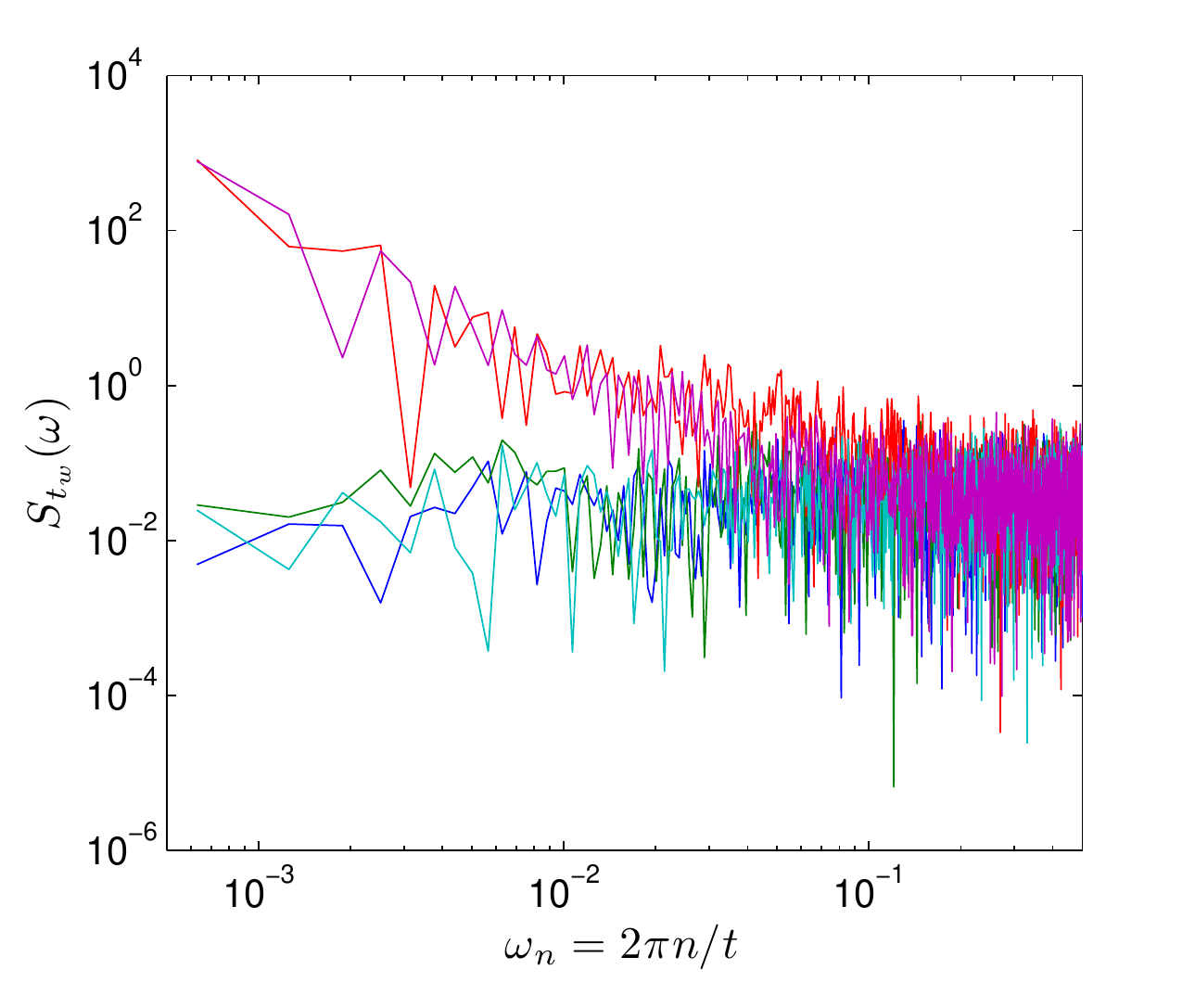}
 		\caption{The corresponding spectra of five blinking quantum dots. The two particles' spectra types are clearly observed; Trapped particles exhibit white noise (green, blue, cyan) versus the non-trapped particles with $1/f^{\beta}$ noise (red, pink). Notice that differentiation between the two population is manifest in sufficient low frequencies, where in higher frequencies all the spectra are observed in a similar order of magnitude. For that reasons, detecting the non-frozen particles needs long measurement time (corresponds to low frequencies).        }
 	\label{fig:BQD.White}
 \end{figure}

 \begin{figure}[t!]
 	\centering
 		\includegraphics[trim=0 0 0 00 , width=\columnwidth]{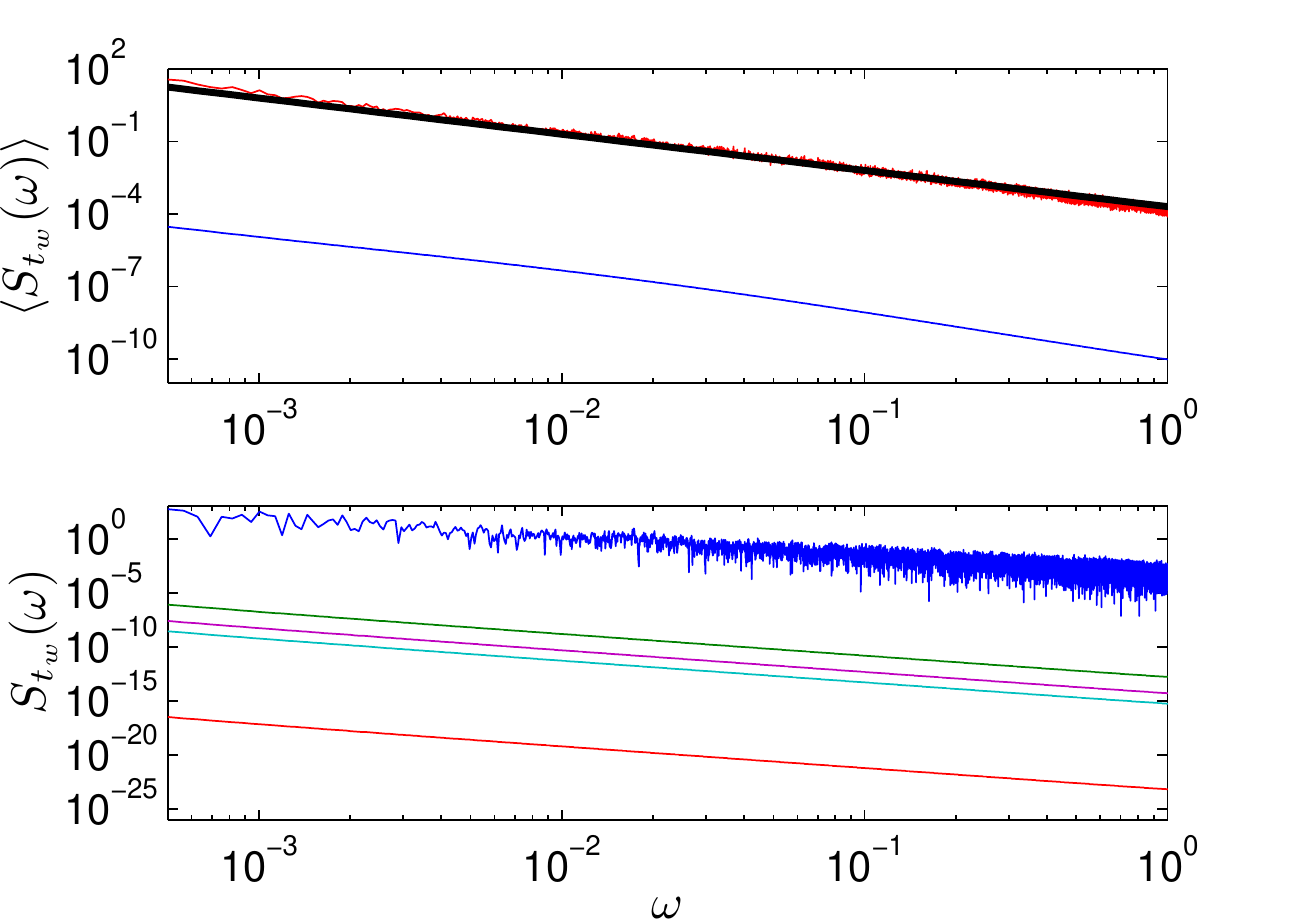}
 		\caption{The spectrum for the deterministic intermittent map with $\alpha=0.5$, waiting time $t_w=10^6$ and measurement time $t=10^5$. In the upper panel we present the single-particles measurements (red line) and its $1/f^{\beta}$ noise prediction Eq.~\eqref{eq06} (black curve). The blue line represents the average spectrum of the trapped particles (i.e. the particles that do not cross the threshold).  In the lower panel five realizations of the spectrum are presented. The trapped particles (four bottom realizations) exhibit significantly low power related to the non-trapped particle (upper blue curve).   }
 	\label{fig:untitled} \end{figure}

As was mentioned, a constant signal in the time interval $(t_w,t_w+t)$  has no contribution to the measured spectrum in natural frequencies since
\begin{equation}
\left|\int_{t_w}^{t_w+t}e^{\imath\omega t'}{\rm d}t'\right|^2=\left|\frac{1-\cos(\omega t)}{2\imath\omega}\right|^2_{\omega=2\pi n/t}=0.
\end{equation}
In that case the trapped particles' spectrum is zero, and thus cannot be detected.
Addition of white (thermal) noise into the process, $\langle S(\omega)\rangle_{\rm th}=\sigma^2$, is reflecting in the spectrum as 
\begin{equation}
{\cal S}_{t_w}(\omega)^{\rm new}_{\rm mac}={\cal S}_{t_w}(\omega)_{\rm mac}+N\langle S(\omega)\rangle _{\rm th},
\end{equation}
where $N$ is the number of molecules.
The trapped particles generate white noise only, while the non-trapped particles exhibit $1/f^{\alpha}$ noise. Hence, one can distinguish between the two sets by their frequencies-dependent spectra; constant spectra for the trapped ones, versus power-law decay for the non-trapped. We illustrate this idea by presenting raw simulated data of five realizations of the blinking quantum dot model with additive white Gaussian noise with zero mean and variance $\sigma^2=0.05$, with $\alpha=0.5$, $t_w=10^5$ and $t=10^4$ (see Fig.~\ref{fig:BQD.White}). The spectrum was calculated with the standard Matlab FFT function. In that case we sort the particles into two types; trapped and non-trapped without a knowledge whether the particle crossed the threshold or did not. In fact this concept of using the spectrum's frequencies dependence may apply on blinking quantum dot with strong white noise (i.e large $\sigma^2$).  

\subsubsection{Deterministic Intermittent Map}
Any measurement device have a minimal detection power, hence particles with lower power are undetectable. In the deterministic map model we fine that a particle whose signal $x_t$ does not cross the threshold $I^*=0.5$ exhibits very low power. A justification to this statement is illustrated in Fig.~\ref{fig:untitled}, where clearly the trapped particles exhibit very low power (related to the non-trapped particles). We present (bottom panel in Fig.~\ref{fig:untitled}) five realizations of the spectrum. In that case, where the minimal detection power is around $10^{-5}$, realizations with lower power are undetectable.

\hspace{1mm}

\subsubsection{Classification}
Following App.~\ref{OUclassification} and the current appendix, we claim that distinguishing between two particles' populations; trapped and non-trapped, may not be trivial and based on two main aspects; The first is the magnitude comparison of the spectra, where a trapped particle exhibits small
amplitude in comparison to the non-trapped particles. In that case the sensitivity threshold of the measurement device effectively determines which particle is
detectable, i.e. if its corresponding power is smaller
than a certain value it cannot be detected. The second
aspect of the differentiation between particles' set is the spectrum frequency dependence. For example we assume each particle has (thermal) white noise in
addition. In that case a localized particle may produce a power in the same order of magnitude as the non-trapped power. However, the trapped particle exhibits only the
white noise, while a non-trapped particle provides $1/f^{\alpha}$ noise. The latter gives larger power in lower frequencies while the white noise spectrum is simply a constant.
Therefore we find that two criteria to distinguish between the two sets; the magnitude and the frequency dependence of the spectrum, both eventually present the two sided of the same coin.

\bibliography{./bibliography}

\end{document}